\documentclass[iop,tighten]{emulateapj}

\usepackage{graphicx}
\usepackage{amssymb}
\usepackage{amsmath}
\usepackage{gensymb}
\usepackage{apjfonts}
\usepackage{natbib}
\usepackage{times}
\usepackage{microtype}

\newcommand{\chandra}{\textit{Chandra }}
\newcommand{\hst}{\textit{HST }}
\newcommand{\iue}{\textit{IUE }}

\newcommand{\einstein}{\textit{Einstein }}
\newcommand{\rosat}{\textit{ROSAT }}
\newcommand{\xmm}{\textit{XMM-Newton }}

\newcommand{\nustar}{\textit{NuSTAR }}
\newcommand{\gmm}{$\Gamma$ }
\newcommand{\chisq}{$\chi\sp{2}$ } 
\newcommand{\xspec}{\textsc{xspec} }
\newcommand{\nh}{$N\sb{\rm H}$ }
\newcommand{\nhgal}{$N\sb{\rm H}\sp{\it Gal}$ }
\newcommand{\nhunits}{$\times 10\sp{22}\;\textrm{cm}\sp{-2}$}
\newcommand{\nhshort}{$\textrm{cm}\sp{-2}$}
\newcommand{\logxi}{$\textrm{log}\;\xi$}
\newcommand{\lum}{{\rm erg\,s^{-1}}}
\newcommand{\flux}{{\rm erg\,s^{-1}cm^{-2}}}
\newcommand{\kms}{{\rm km\,s^{-1}}}
\newcommand{\ang}{\text{\normalfont\AA}}
\newcommand{\ciao}{\textsc{ciao} }
\newcommand{\ka}{K$\alpha$ }
\newcommand{\ovi}{\ion{O}{6}}
\newcommand{\nv}{\ion{N}{5}}
\newcommand{\civ}{\ion{C}{4}}
\newcommand{\siiv}{\ion{Si}{4}}
\newcommand{\aox}{$\alpha\sb{\rm OX}$}
\newcommand{\luv}{${\rm log}~l\sb{\rm 2500\ang}$}
\newcommand{\pgten}{PG~1004+130}
\newcommand{\rg}{$R\sb{g}$}

\begin{document}

\submitted{Accepted 2015 May 5}

\shorttitle{Scott et al.}
\shortauthors{Scott et al.}

\title{Ultraviolet/X-ray variability and the extended X-ray emission of the \\
radio-loud broad absorption line quasar PG~1004+130}
 
\author{A.~E.~Scott\altaffilmark{1,2}, 
  W.~N.~Brandt\altaffilmark{1,2,3},
  B.~P.~Miller\altaffilmark{4},
  B.~Luo\altaffilmark{1,2},
  S.~C.~Gallagher\altaffilmark{5,6}}

\affil{
  $^{1}$Department of Astronomy \& Astrophysics, 525 Davey Laboratory, Pennsylvania State University, University Park, PA 16802, USA; amyscott@psu.edu\\
  $^{2}$Institute for Gravitation and the Cosmos, Pennsylvania State University, University Park, PA 16802, USA\\
  $^{3}$Department of Physics, 104 Davey Laboratory, Pennsylvania State University, University Park, PA 16802, USA\\
  $^{4}$Department of Chemistry and Physical Sciences, The College of St. Scholastica, Duluth, MN 55811, USA\\
  $^{5}$Department of Physics and Astronomy, University of Western Ontario, London, ON, N6A 3K7, Canada\\
  $^{6}$Visiting Fellow, Yale Center for Astronomy and Astrophysics, Yale University, P.O. Box 208120, New Haven, CT 06520, USA  
}

\begin{abstract}
We present the results of recent \textit{Chandra}, \textit{XMM-Newton}, and \textit{Hubble Space Telescope} observations of the radio-loud (RL), broad absorption line (BAL) quasar PG~1004+130.  We compare our new observations to archival \mbox{X-ray} and UV data, creating the most comprehensive, high signal-to-noise, multi-epoch, spectral monitoring campaign of a RL BAL quasar to date.  We probe for variability of the \mbox{X-ray} absorption, the UV BAL, and the \mbox{X-ray} jet, on month--year timescales.  The \mbox{X-ray} absorber has a low column density of $N\sb{\rm H}=8\times10\sp{20}-4\times10\sp{21}$~\nhshort\/ when it is assumed to be fully covering the \mbox{X-ray} emitting region, and its properties do not vary significantly between the 4 observations.  This suggests the observed absorption is not related to the typical ``shielding gas'' commonly invoked in BAL quasar models, but is likely due to material further from the central black hole.  In contrast, the \civ\/ BAL shows strong variability.  The equivalent width (EW) in 2014 is $EW=11.24\pm0.56~\ang$, showing a fractional increase of $\Delta EW / \langle EW \rangle=1.16\pm0.11$ from the 2003 observation, 3183~days earlier in the rest-frame.  This places \pgten\/ among the most highly variable BAL quasars.  By combining \chandra observations we create an exposure $2.5\times$ deeper than studied previously, with which to investigate the nature of the \mbox{X-ray} jet and extended diffuse \mbox{X-ray} emission.  An \mbox{X-ray} knot, likely with a synchrotron origin, is detected in the radio jet $\sim8\arcsec$ (30~kpc) from the central \mbox{X-ray} source with a spatial extent of $\sim4\arcsec$ (15~kpc).  No similar \mbox{X-ray} counterpart to the counterjet is detected.  Asymmetric, non-thermal diffuse \mbox{X-ray} emission, likely due to inverse Compton scattering of Cosmic Microwave Background photons, is also detected.\\
\end{abstract}


\keywords{galaxies: active -- quasars: individual (PG~1004+130) -- quasars: absorption lines -- X-rays: galaxies}


\section{Introduction}
\label{section:intro}
Active Galactic Nuclei (AGN) can significantly affect the properties of their host galaxies through 
mechanical feedback provided by winds and jets.  Fast ($v\sim0.01-0.1c$) outflowing winds manifest 
themselves observationally as blueshifted, broad absorption lines (BALs) in the UV/optical spectra 
of $\sim15$\% of quasars (e.g.,~\citealt{hewett03}), although it is likely that they exist in most 
quasars and we only detect them when our line-of-sight passes directly through the outflowing material 
(e.g.,~\citealt{weymann91}).  As this material is not static, variability is expected in the BALs on a range of 
timescales characteristic of, e.g.,~variations in the ionization parameter and/or changes in the
flow structure or disk rotation.  There have been many studies of the UV variability of BALs in samples of quasars 
(e.g.,~\citealt{barlow93,lundgren07,gibson08,gibson10,capellupo11,capellupo12,capellupo13,filizak12,filizak13,wildy14,welling14}).
These find that BAL variability is both more common and stronger on 
longer rest-frame timescales and in weaker and higher velocity BALs, 
with the dominant parameter being the equivalent width (EW) of the BAL \citep{filizak13}.

In the disk-wind scenario, the wind is launched from the accretion disk at $\sim1000$\,\rg\, and is 
driven by UV line pressure.  For a central source emitting typical levels of \mbox{X-rays}, this 
requires some ``shielding gas'' to be located between it and the wind to prevent over-ionization of 
the outflow (e.g.,~\citealt{murray95,proga00}).  This gas may be responsible for the absorption commonly 
observed in the \mbox{X-ray} spectra of BAL quasars which is well fit with either a partially covering 
and/or a partially ionized absorption model with typical column densities of 
\nh$\sim10\sp{21-23}~\textrm{cm}\sp{-2}$ (e.g.,~\citealt{gallagher02,giustini08,fan09,streblyanska10}).  
This \mbox{X-ray} absorption is also observed to vary in some cases \citep{gallagher04,giustini11,saez12}, 
possibly as a result of complex dynamics related to rotating, infalling, or outflowing material.  Compared 
with non-BAL quasars of a similar UV luminosity, BAL quasars are underluminous in \mbox{X-rays} 
(e.g.,~\citealt{green95,green96,gallagher99,gallagher06,brandt00}), but, after correcting for the effects 
of absorption, a majority are found to have similar underlying \mbox{X-ray} spectral properties to those of 
non-BAL quasars (e.g.,~\citealt{gallagher02,page05}).  However, some BAL quasars are likely to be 
intrinsically \mbox{X-ray} weak \citep{teng14,luo14} and in such cases, shielding gas is not required 
to ensure launching of the wind. 

Approximately 10\% of AGN have powerful, highly collimated and relativistic jets which emerge from the
inner regions ($\sim10$\,\rg) of the AGN and emit strongly at radio and \mbox{X-ray} frequencies via
synchrotron and Synchrotron Self Compton (SSC;~\citealt{band86}) processes.  Radio-loud quasars (RLQs)
which possess such jets have higher \mbox{X-ray} luminosities than radio-quiet quasars 
(RQQs;~e.g.,~\citealt{zamorani81,worrall87,miller11,scott11}),
and flatter (harder) \mbox{X-ray} power-law spectral indices (e.g.,~\citealt{reeves00,saez11}) as a result of
this additional jet-linked \mbox{X-ray} emission component \citep{sambruna99}.  \mbox{X-ray} counterparts
to extended radio jets have been observed in many AGN (e.g.,~\citealt{sambruna04,worrall09,marshall11}).
Diffuse \mbox{X-ray} emission has also been detected tracing the direction of radio jets and unseen
counterjets (e.g.,~\citealt{schwartz06}) and radio lobes (e.g.,~\citealt{croston05}) and is likely due
to inverse Compton scattering of Cosmic Microwave Background (CMB) photons (IC/CMB) by relativistic jet electrons.

It was originally thought that quasars could not host both a BAL wind and a radio jet (e.g.,~\citealt{stocke92}).
However, many subsequent discoveries of radio-loud (RL) BAL quasars have now been made (e.g.,~\citealt{becker97,becker00,brotherton98,menou01}) 
although BALs are still not found in quasars with the highest levels of radio-loudness (e.g.,~\citealt{gregg06,shankar08}).  
This suggests that the radio jet and BAL wind may affect each other, despite being emitted from different
locations within the AGN.  It is perhaps more likely that physical properties of the central engine such
as mass accretion rate and the spin of the black hole are responsible for the presence of either phenomenon,
although no significant differences in the black hole masses or the Eddington ratios of radio-quiet (RQ) and
RL BAL quasars have been found \citep{bruni14}.  RL and RQ BAL quasars also show similar optical properties
\citep{runnoe13,rochais14}.  However \citet{baskin15} suggest that differences in the BAL properties of RQ
and RL quasars are due to a softer UV/optical SED, indicated by a lower \ion{He}{2} EW, which are preferentially
found in RL BALs.  BAL variability in RLQs does not depend strongly on radio loudness \citep{filizak13},
but lobe-dominated RLQs may show greater fractional BAL variability than core-dominated RLQs \citep{welling14}.
When considering variability on the same timescales, RLQs show a lower fractional variability than RQQs,
$\sim40\pm20$\% \citep{welling14}.  RL BAL quasars are \mbox{X-ray} weak compared to RL non-BAL quasars as
expected, but to a lesser degree than the difference between RQ BAL and RQ non-BAL quasars.  This suggests that
the \mbox{X-ray} absorption does not affect all of the jet-linked \mbox{X-ray} emission \citep{miller09}.

\pgten\/ (also known as PKS~1004+13 and 4C~13.41) is one of the best-studied examples of a RL BAL quasar.
It is optically bright ($B=15.8$;~\citealt{veron10}), low redshift ($z=0.2406$,~\citealt{eracleous04}),
and has a black hole mass of $M\sb{\rm BH}=1.87\sp{+0.36}\sb{-0.45}\times10\sp{9}M\sb{\odot}$ \citep{vestergaard06}.
\textit{International Ultraviolet Explorer} (\textit{IUE}) and \textit{Hubble Space Telescope} (\textit{HST})
spectra of \pgten\/ show blueshifted absorption ($v\sim10,000~\kms$) in the high-ionization lines \ovi, \nv, \siiv,
and \civ\/ \citep{wills99}, and it has a \civ\/ BALnicity 
index\footnote{${\rm BI}=-\displaystyle\int_{25000}^{3000}\nolimits\left[1-\frac{f(v)}{0.9}\right]dv$\vspace*{0.3cm}\\where 
$f(v)$ is the normalized flux as a function of velocity displacement from the rest-frame wavelength of the emission 
line.  The BI does not include the first $2000~\kms$ of the absorption trough \citep{weymann91}.} of 
${\rm BI}\sim850\;\kms$ \citep{wills99}.  It also shows strong \civ\/ BAL variability, 
with large fractional changes in EW \citep{welling14}.  

Like most BAL quasars, \pgten\/ is \mbox{X-ray} weak and was undetected in an early \einstein observation
\citep{elvis84}.  The first \mbox{X-ray} detection and spectra of \pgten\/ were obtained with \xmm in 2003
and \chandra in 2005, and were reported in \citeauthor{miller06}~(\citeyear{miller06}; hereafter M06).
Although the \xmm spectrum did not show any significant \mbox{X-ray} absorption, the \chandra spectrum
obtained 20-months (observed-frame) later, revealed complex  \mbox{X-ray} absorption best-fit with a
partial-covering model (\nh$=1.2\sp{+0.83}\sb{-0.84}$\nhunits, $f=0.49\sp{+0.14}\sb{-0.26}$,
\gmm$=1.37\sp{+0.18}\sb{-0.22}$).  \pgten\/ has also been observed by the \textit{Nuclear Spectroscopic
  Telescope ARray} (\textit{NuSTAR}; \citealt{Harrison13}).  It was weakly detected in the $10-20$~keV band,
suggesting that the shielding \mbox{X-ray} gas is either Compton thick or that \pgten\/ is intrinsically
\mbox{X-ray} weak, although neither scenario was strongly preferred over the other \citep{luo13}.  If the
absorbing material is Compton thick, we would expect to detect a strong Fe~\ka emission line, but
this is not observed (M06;~\citealt{luo13}).  Its absence could be due to \mbox{X-ray} emission related
to a jet diluting the spectrum \citep{luo13}, a scenario also favored by broadband modeling \citep{kunert15}.  

\pgten\/ is RL with a radio
loudness parameter of $R\sp{*}=209$ (\citealt{wills99}; where $R\sp{*} = f\sb{\rm 5\,GHz}/f\sb{\rm 2500\ang}$,
and $R\sp{*}>10$ indicates a RLQ) and its extended radio emission is of a HYMOR (HYbrid MORphology;~\citealt{gopal00})
nature.  The surface-brightness of the radio lobe to the SE decreases outwards i.e.,~a Fanaroff-Riley I (FRI)
classification \citep{fanaroff74}, while the surface-brightness of the NW lobe increases outwards (FRII).  This
asymmetry is likely due to twin jets propagating into dissimilar large-scale environments, i.e.,~the density of
the surrounding medium is higher to the SE so that the jet is more quickly decollimated \citep{gopal00}.  The
\mbox{X-ray} imaging analysis reported in M06 revealed an \mbox{X-ray} counterpart to the radio jet extending
to the SE $\sim8\arcsec$ from the nucleus and upstream from the peak of the radio jet emission.  Large-scale
diffuse \mbox{X-ray} emission was also detected, extending $40-50\arcsec$ outward from the nucleus with a flux
of $\sim(4.2-4.5)\times10\sp{-15}~\flux$ (assuming a power-law model with \gmm$=1.8$; M06).  

In this work, we present results from newly obtained \mbox{X-ray} and UV observations of \pgten, and compare
them to archival data.  This allows us to probe for variability of the \mbox{X-ray} absorption, the \civ\/
broad absorption line, and the \mbox{X-ray} jet emission on month-to-year timescales.  By combining multiple
\chandra observations we obtain the deepest imaging to date of an \mbox{X-ray} jet in a HYMOR radio source,
with $2.5\times$ greater sensitivity than the original data used in M06.  This allows us to place further
constraints on the spatial and spectral properties of the \mbox{X-ray} jet, and larger-scale diffuse \mbox{X-ray}
emission.  This study of \pgten\/ represents the most comprehensive, high signal-to-noise (S/N), multi-epoch,
\mbox{X-ray} and UV spectral monitoring campaign of a RL BAL quasar, with results applicable to this class of
quasars as a whole.  A description of our new and archival \mbox{X-ray} and UV observations, and details of
the data reduction are given in Section~\ref{section:obs}.  In Section~\ref{section:var} we present an analysis
of the \mbox{X-ray} and UV emission from the nuclear source, including the \mbox{X-ray} absorption properties
and the \civ\/ BAL variability.  In Section~\ref{section:image} we present analysis of the extra-nuclear
\mbox{X-ray} jet and diffuse emission.  We discuss our results in Section~\ref{section:disc} and summarize
our conclusions in Section~\ref{section:sum}.  A standard cosmology in which
$H\sb{0}=69.7\;\textrm{km}\,\textrm{s}\sp{-1}\,\textrm{Mpc}\sp{-1}$, $\Omega\sb{\Lambda}=0.7185$, and
$\Omega\sb{\rm M}=0.2815$ is assumed throughout \citep{hinshaw13}.  This gives a luminosity distance to
\pgten\/ of $d\sb{\rm L}=1216$~Mpc, and hence an angular scale of 3.8~kpc per arcsec.  Errors quoted on
X-ray quantities are 90\%, unless otherwise stated.


\section{Observations}
\label{section:obs}
\subsection{\chandra Data Reduction}
\label{section:chandra}
\pgten\, has been observed twice with the Advanced CCD Imaging Spectrometer (ACIS; \citealt{acis}) on board 
\chandra \citep{chandra}.  We present results from both a recently obtained 60~ks observation (taken in 
2014 Mar) and a 40~ks observation taken in 2005 Jan which was previously reported in M06.  Details of both
observations are given in Table~\ref{table:obslog}.  We reduce each data set in the same, consistent 
way using \ciao version 4.6.1 \citep{ciao} and version 4.6.2 of the CALDB.  The \texttt{chandra\_repro} 
script is used to apply the latest calibration data.  Neither observation suffers from significant background 
flaring, but we use a standard $3\sigma$ clipping of outliers from the $2.4-6$~keV light curve to select 
good-time intervals (GTIs) where the count rate was $\le0.23\,{\rm s}\sp{-1}$ or $\le0.21\,{\rm s}\sp{-1}$ 
in 2005 and 2014, respectively.  Improved screening for potential cosmic-ray background events was carried 
out on the 2014 data which were taken in VFAINT mode.  Neither observation suffers from pile-up in the nuclear 
source.  We use \texttt{specextract} to obtain spectra for the point-like nucleus using a circular extraction 
region with a 4\arcsec\/ radius centered on the X-ray source position, and background spectra are extracted
from a 16\arcsec\/ radius circle, offset from the source position.  These spectra are binned to a minimum of
20 counts per bin to ensure the validity of \chisq statistics during the subsequent spectral fitting.  Auxiliary
response files are generated using the latest calibrations to include the time-dependent quantum-efficiency
(QE) degradation due to contamination build-up on the ACIS optical-blocking filter.  \mbox{X-ray} spectra of
the jet emission are also extracted from a rectangular region of size $6.4\arcsec \times 4.3\arcsec$, offset 
by 8\arcsec\/ from the nuclear source, and at a position angle\footnote{All angles are measured from North, 
through East.} of 130\degree\/ from the source position in both observations (see Section~\ref{section:image}).  
Background spectra are extracted from the same 16\arcsec\/ radius circular regions as used previously, but in 
this case no binning is applied, and the spectra are modeled in \xspec using the Cash statistic \citep{cash79} 
due to the low number of counts in the spectra.\\ 

\begin{table*}[ht!]
\centering
\caption{\mbox{X-ray} Observation Log}
\label{table:obslog}
\vspace*{-0.4cm}
   \begin{center}
    \begin{tabular}{llllccc}
      \hline \hline\\[-2.0ex]
      \multicolumn{1}{l}{Start Time} &
      \multicolumn{1}{l}{Obsid} &
      \multicolumn{1}{l}{Observatory} &
      \multicolumn{1}{l}{Detector(s)} &
      \multicolumn{1}{c}{Exposure Time} &
      \multicolumn{1}{c}{Nuclear Source Counts$^{a}$} &
      \multicolumn{1}{c}{Observed $0.5-8$~keV Flux$^{b}$}\\
      \multicolumn{1}{l}{(UTD)} &
      \multicolumn{1}{l}{ } &
      \multicolumn{1}{l}{ } &
      \multicolumn{1}{l}{ } &
      \multicolumn{1}{c}{(ks)} &
      \multicolumn{1}{c}{ } &
      \multicolumn{1}{c}{($\times 10\sp{-13}\flux$)} \\ [0.5ex]
      \hline\\ [-2.0ex]
      2003 May 04 19:06:59  & 0140550601  & \xmm      & EPIC-MOS  &  21.51  & $1171\sp{+35}\sb{-34}$  & $3.49\sp{+0.17}\sb{-0.23}$  \\ 
                            &             &           & EPIC-pn   &  18.08  & $1574\sp{+41}\sb{-40}$  & \ldots                      \\ [1.0ex]
      2005 Jan 05 16:44:27  & 5606        & \chandra  & ACIS-S    &  39.29  & $1763\sp{+43}\sb{-42}$  & $4.87\sp{+0.29}\sb{-0.30}$  \\ [1.0ex]
      2013 Nov 05 05:58:10  & 0728980201  & \xmm      & EPIC-MOS  &  63.14  & $2399\sp{+50}\sb{-49}$  & $2.78\sp{+0.10}\sb{-0.10}$  \\ 
                            &             &           & EPIC-pn   &  54.99  & $3598\sp{+61}\sb{-60}$  & \ldots                      \\ [1.0ex]
      2014 Mar 20 10:42:47  & 16034       & \chandra  & ACIS-S    &  60.97  & $1599\sp{+41}\sb{-40}$  & $3.34\sp{+0.22}\sb{-0.21}$  \\ [1.5ex]
      \hline
    \end{tabular}
    \end{center}
   \vspace*{-0.2cm}
    \begin{raggedright}
    \hspace*{0.6cm}\textbf{Notes.} \\
    \hspace*{0.6cm}$^{a}$The number of observed counts in the nuclear source extraction region.  For \textit{XMM-Newton} this includes counts 
                   in the energy range $0.5-10$~keV, from 
    \hspace*{0.6cm}a circular extraction region with a radius of 24.6\arcsec\/; for \textit{Chandra} this includes counts in the range $0.5-8$~keV 
                   from a 4\arcsec\/ radius region.\\
    \hspace*{0.6cm}$^{b}$The observed $0.5-8$~keV fluxes are determined from an intrinsically absorbed power-law model.  For the \xmm observations the MOS and pn\\
    \hspace*{0.6cm}spectra are modeled simultaneously giving a single flux estimate.\\
    \vspace*{0.2cm}
    \end{raggedright}
\end{table*}


\subsection{\xmm Data Reduction}
\label{section:xmm}
We also consider two \xmm \citep{xmm} European Photon Imaging Camera (EPIC;~\citealt{MOS,pn}) observations
of \pgten; a more recent 60~ks observation taken in 2013 Nov as part of AO-12 and a 20~ks archival observation
from 2003 May, first reported in M06.  Neither of these observations were affected by substantial background
flaring, therefore GTIs were defined as times where the count rate was $\le0.4\,{\rm s}\sp{-1}$ in the pn
detector and $\le0.35\,{\rm s}\sp{-1}$ in the MOS detectors (although for the 2013 observation this was
reduced to $0.115\,{\rm s}\sp{-1}$ to exclude 2 short flares).  Final exposure times and details of the
observations are given in Table~\ref{table:obslog}.  Data from each observation were reduced in a consistent
and standard way using \textsc{sas} version 13.5.0.  Spectra of the point-like nucleus were extracted from
a circular region with a radius of 24.6\arcsec, as in M06, centered on the source position.  Neither
observation suffers from pile-up of the nuclear source.  Circular regions on the same CCD but offset from
the source with radii of 43\arcsec\/ and 34\arcsec\/ were used to extract background spectra from the MOS
and pn detectors, respectively.  Auxiliary response files and redistribution matrices were generated using
\texttt{arfgen} and \texttt{rmfgen}.  The resulting spectra from the two MOS detectors were combined, and
both the resulting MOS and pn spectra are binned to a minimum of 20 counts per bin using the Ftool
\texttt{grppha}.\footnote{http://heasarc.gsfc.nasa.gov/docs/software/ftools/}\\


\subsection{\hst Data Reduction}
\label{section:hst}
We used the Cosmic Origins Spectrograph (COS;~\citealt{cos}) onboard the \hst to obtain a current-epoch
Near-UV (NUV) spectrum covering $\lambda\sb{\rm obs}=1670-2120$~\AA\/ which includes the \ion{C}{4} BAL 
region of \pgten.  A 2562~s exposure was obtained with 
the G230L grating at a central wavelength of 2950~\AA\, on 2014 Jan 20.  The G230L grating was selected
to provide spectral resolution exceeding archival UV observations while spanning the entire \ion{C}{4}
emission line and BAL absorption region on the A stripe.  The target was centered in the cross-dispersion
and along-dispersion directions using an ACQ/PEAKXD and ACQ/PEAKD acquisition sequence.  Four focal plane
grating offset positions (FPPOS=1--4) were employed to accumulate the spectrum on different regions of the
detector.  Data were taken in the TIME-TAG mode to retain the arrival time information for each event.
Standard pipeline\footnote{http://www.stsci.edu/hst/cos/documents/handbooks/datahandbook/} processing steps
were applied, including Doppler correction for the shift induced by the orbital motion of \textit{HST},
non-linearity correction for detector deadtimes at high count rates, flat-field correction for pixel
sensitivities, wavelength calibration with a flashed line-lamp, bad pixel flagging, extraction of the
background-subtracted one-dimensional spectrum, conversion from counts to flux, and conversion to
heliocentric wavelengths.  The final spectrum has a S/N ratio of $\sim10$ across the \ion{C}{4} BAL.
No significant temporal or spectral variability is present over the duration of the observation.\\


\section{X-ray and UV absorption variability}
\label{section:var}
\subsection{Long-term Light Curve}
\label{section:lightcurve}
\begin{figure}
  \centering 
  \hspace*{-0.25cm}
    \includegraphics[width=0.51\textwidth]{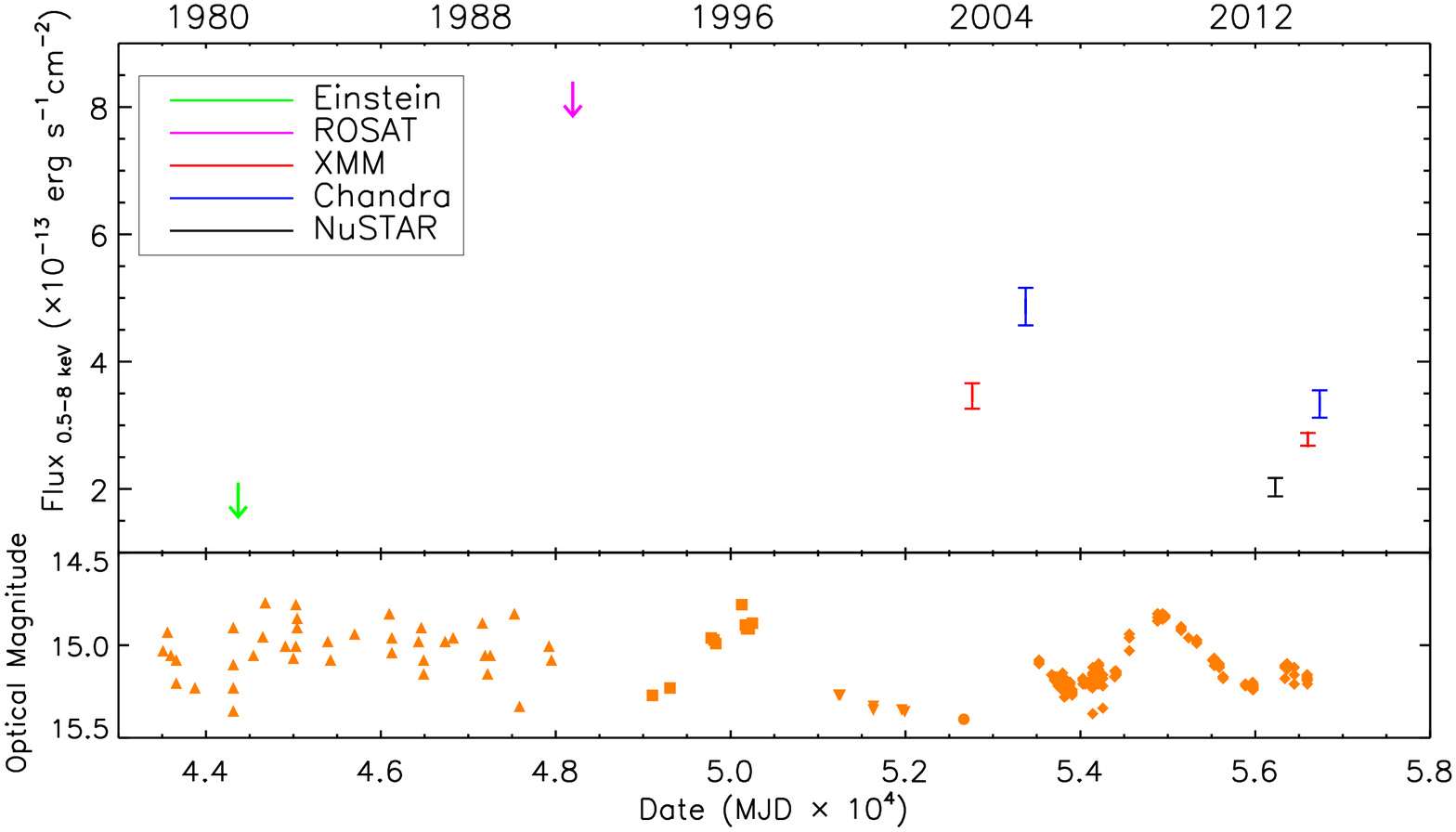}
  \caption{Long-term light curves of \pgten.  Top -- the observed
    $0.5-8$~keV fluxes estimated from an absorbed power-law model.
    Fluxes from both \xmm observations considered in this work
    are shown in red, and the fluxes from both \chandra observations
    are shown in blue.  Each is plotted with a 90\% error bar.  The
    observed $0.5-8$~keV flux extrapolated from a \nustar observation
    is shown in black \citep{luo13}.  The green arrow represents a
    $3\sigma$ upper limit from an early \einstein observation
    \citep{elvis84}, and the magenta arrow represents an upper limit
    from the \rosat All Sky Survey \citep{voges99}.  Bottom --
    Filled orange symbols show optical magnitudes from
    \citeauthor{smith93}~(\citeyear{smith93}; upwards triangles),
    \citeauthor{garcia99}~(\citeyear{garcia99}; squares), 
    \citeauthor{stalin04}~(\citeyear{stalin04}; downwards triangles),
    SDSS (\citealt{sdss}; circle), and the 
    Catalina Real-Time Transient Survey (\citealt{drake09}; diamonds).\\}
  \label{fig:lc}
\end{figure}

Figure~\ref{fig:lc} shows long-term light curves for \pgten. The top panel shows the observed $0.5-8$~keV 
\mbox{X-ray} fluxes (not corrected for Galactic or intrinsic $N\sb{\rm H}$), estimated from an absorbed 
power-law model in which the power-law slope is free to vary.  \xmm fluxes are shown in red, and \chandra 
fluxes are shown in blue, both with 90\% errors.  Shown in black is the flux extrapolated from a \nustar 
observation in 2012 Oct \citep{luo13}, the green arrow indicates a $3\sigma$ upper limit from an early 
\einstein observation \citep{elvis84}, and the magenta arrow is an upper limit derived from the \rosat 
All Sky Survey \citep{voges99}.  Each was extrapolated to a $0.5-8$~keV flux by assuming a power-law 
slope of $\Gamma=1.5$, typical of radio-loud AGN (e.g.,~\citealt{reeves00}).  

The bottom panel shows historical monitoring giving an indication of the typical optical variability
($\sim0.5$~mag) of \pgten.  $V$--band magnitudes are estimated from the unfiltered $P$ magnitudes given
in \citet{smith93}, differential $V$--band photometry given in \citet{garcia99}, differential $R$--band
photometry given in \citet{stalin04}, photometry from SDSS \citep{sdss}, and unfiltered CCD magnitudes
from the Catalina Real-Time Transient Survey \citep{drake09}. 

There exists a well-known correlation between the \mbox{X-ray}-to-optical power-law slope parameter, 
\aox\footnote{$\alpha\sb{\rm OX}=0.38\,\textrm{log}\left(f\sb{\rm X}/f\sb{\rm O}\right)$ where 
$f\sb{\rm X}$ and $f\sb{\rm O}$ are the rest-frame flux densities at $2\,\textrm{keV}$ and $2500\,\textrm{\AA}$, respectively.}
\citep{tananbaum79}, and the rest-frame UV luminosity, due to the physical interplay between the UV 
disk photons and the \mbox{X-ray} corona (e.g.,~\citealt{vignali03,strateva05,steffen06,young10,lusso10}).  
Adopting \luv$=30.78$ from \citet{shen11}, and after correcting for the excess \mbox{X-ray} luminosity 
expected in RLQ \citep{miller11}, an \mbox{X-ray} flux of $F\sb{\rm 0.5-8~keV}=(75-95)\times10\sp{-13}~\flux$ 
is expected.  This is $15-35\times$ higher than we observe in any of the 4 \xmm and \chandra observations showing 
the persistent \mbox{X-ray} weakness of \pgten.

Flux estimates of the same object measured by the EPIC and ACIS detectors are known to show discrepancies 
with ACIS measuring fluxes that are typically $\sim 10-15\%$ higher (e.g.,~\citealt{nevalainen10,tsujimoto11}).
The 2005 ACIS flux is $\sim40$\% higher than the 2003 flux measured by EPIC, and the 2014 ACIS flux is 
$\sim20$\% higher than the 2013 EPIC flux.  These differences, which are larger than those typically expected, 
are suggestive of a real flux change in the source between the observations, rather than simply calibration 
differences.  When we consider the two \textit{XMM-Newton}, or the two \chandra observations separately, which 
reduces calibration effects, we see a significant\footnote{$4.5\sigma$ for \xmm and $6.7\sigma$ for 
\textit{Chandra}.} flux decrease on timescales of $\sim8$~years in the rest-frame of \pgten.  A similarly 
significant decrease is observed in the hard band flux ($2-8$~keV).  The lower flux measured by \nustar is also 
suggestive of a real flux change in the source.\\

\subsection{X-ray Spectral Fitting of the Nuclear Source}
\label{section:fitting}
The grouped \mbox{X-ray} spectra were fit with physically motivated models using \xspec version 12.8
\citep{xspec}.  Each spectral model includes absorption fixed at the value of the Galactic component
in the direction of the source.  This is estimated using equation 7 from \citet{willingale13},
$N\sb{\rm H\textsc{i}}=3.70\times10\sp{20}\;\textrm{cm}\sp{-2}$ from \citet{dickey90}, and
$E(B-V)=0.0331\;\textrm{mag}$ from \citet{schlafly11}, resulting in \nhgal$=4.1\times 10\sp{20}\;\textrm{cm}\sp{-2}$.
Absorption is modeled using \texttt{tbabs} with abundances from \citet{wilms00} and cross-sections
from \citet{vern96}.  Unless stated otherwise, the \xmm spectra were modeled over the energy range
$0.5-10$~keV.  The MOS and pn data were fit simultaneously using the same parameters with a freely
varying constant added to the model to account for calibration offsets between the two detectors
\citep{mateos09}.  The \chandra ACIS spectra were modeled over the energy range $0.5-8$~keV.  

\begin{figure}[h]
  \centering 
    \includegraphics[width=0.48\textwidth]{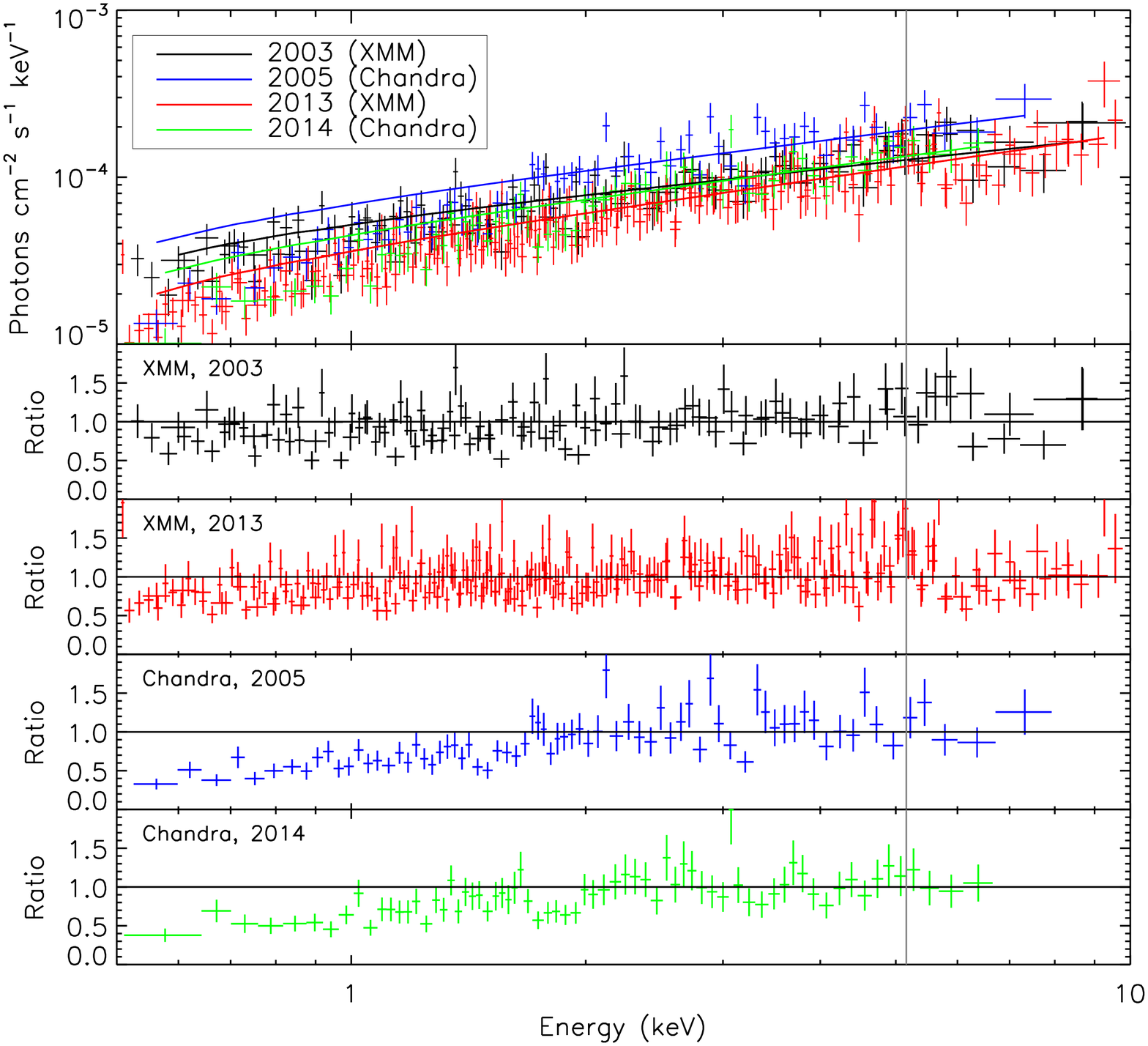}
  \caption{\mbox{X-ray} spectra from each of the 4 epochs. The \xmm MOS and
    pn spectra are fit simultaneously (data from MOS1 and MOS2 are
    combined) and are plotted in the same color on this figure for
    clarity.  Each of the data sets are modeled with a power law (with
    only a contribution from Galactic absorption) over $2-8$~keV
    (\textit{Chandra}) and $2-10$~keV (\textit{XMM-Newton}).  This
    model is shown extrapolated down to $0.5$~keV.  The vertical grey
    line indicates the observed energy at which the neutral iron
    emission line at a rest-frame energy of 6.4~keV is expected.  This
    feature is visible in the 2013 \xmm spectra.  The
    top panel shows the unfolded spectra with $E\times Ef(E)$ plotted
    on the $y$-axis such that a power-law slope of $\Gamma=2$ would be
    horizontal in the figure.  The bottom panels shows the ratio of
    the data to the power-law model, with the residuals at lower
    energies clearly indicating the absorption present in both
    \chandra spectra, and the 2013 \xmm spectra.\\}
  \label{fig:xray_spectra}
\end{figure}

Each of the 4 data sets was first modeled with a simple power-law fit to the high-energy data (above
$2$~keV), absorbed only by a Galactic contribution.  Figure~\ref{fig:xray_spectra} shows the unfolded
spectra (top) and the ratio of the data to this power-law model (bottom).  This clearly shows a flat
power-law slope, as expected for radio-loud AGN such as \pgten.  When the hard-band power-law fit is 
extrapolated down to 0.5~keV, the presence of intrinsic
absorption at lower energies is also clear, and appears to be larger in the \chandra spectra than in
the \xmm spectra.  Such absorption is expected for BAL quasars (e.g.~\citealt{gallagher02}).

\begin{figure}[h]
  \centering 
    \includegraphics[width=0.48\textwidth]{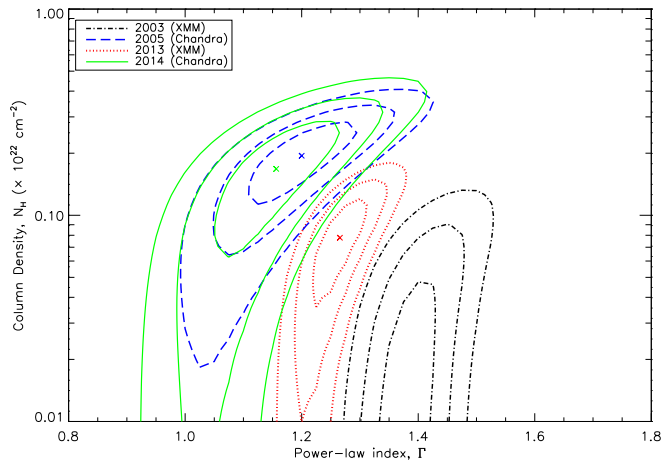}
  \caption{Confidence contours (1, 2 \& 3$\sigma$), for an absorbed
    power-law fit (\texttt{tbabs*ztbabs*po}) for each of the 4
    data-sets.  Crosses indicate the fit parameters which give the
    minimum value of $\chi\sp{2}$.  The \chandra data appears to be
    consistent between each epoch.  The 2003 \xmm data yields
    significantly different spectral parameters; differences that
    appear to be larger than those expected from known calibration
    differences between the detectors.  The 2013 \xmm data also shows
    differences, but of a smaller size, such that they may be
    attributed to calibration effects.\\}
  \label{fig:apo_contours}
\end{figure}

We therefore fit each of the data sets with an absorbed power-law, \xspec model \texttt{tbabs*ztbabs*po}, 
i.e.,~the absorber at the redshift of the source is both fully neutral and fully covering the source.
The best-fitting parameters for each model are listed in Table~\ref{table:fits}.  Confidence contours
(at 1, 2, \& 3$\sigma$) from this absorbed power-law fit are shown in Figure~\ref{fig:apo_contours},
where different colors and linestyles are used for each of the 4 data sets.  Crosses mark the location
of the fit parameters for which \chisq is a minimum, although none is shown for the 2003 \xmm data as
\nh is unconstrained in this fit.  The results from the two \chandra observations (2005, shown in dashed
blue, and 2014, shown in solid green) are highly consistent, showing constrained absorption column densities
of $N\sb{\rm H,\,2005}=(0.19\sp{+0.10}\sb{-0.08})$\,\nhunits\, and 
$N\sb{\rm H,\,2014}=(0.17\sp{+0.13}\sb{-0.12})$\,\nhunits, respectively.  However, 
there are clear differences between these \chandra data and the results from both 
\xmm observations, which show steeper photon indices and lower levels of absorption.  

\begin{table*}[ht!]
\begin{center}
\caption{X-ray Spectral Fitting Results}
    \label{table:fits}
    \begin{tabular}{lllcccccccc}
      \hline \hline\\[-2.0ex]
      \multicolumn{1}{l}{Data} &
      \multicolumn{1}{l}{Model} &
      \multicolumn{1}{c}{\gmm} &
      \multicolumn{1}{c}{\nh} &
      \multicolumn{1}{c}{$f$} &
      \multicolumn{1}{c}{log $\xi$} &
      \multicolumn{1}{c}{H0} &
      \multicolumn{1}{c}{F-test}\\
      \multicolumn{1}{l}{ } &
      \multicolumn{1}{l}{ } &
      \multicolumn{1}{c}{ } &
      \multicolumn{1}{c}{(\nhunits)} &
      \multicolumn{1}{c}{ } &
      \multicolumn{1}{c}{[erg s$\sp{-1}$ cm]} &
      \multicolumn{1}{c}{(\%)} &
      \multicolumn{1}{c}{(\%)}\\[0.5ex]
      \hline\\[-2.0ex]
 \xmm     & Power law                        & $1.37\sp{+0.05}\sb{-0.05}$ &                            &                            &      & 17.3  &       \\[1.0ex]
 2003     & PL + absorption                  & $1.36\sp{+0.07}\sb{-0.05}$ & $\le 0.06$                 &                            &      & 15.7  & 0     \\[1.0ex]
          & PL (2--10 keV)                   & $1.50\sp{+0.11}\sb{-0.14}$ &                            &                            &      & 82.2  &       \\[1.0ex]
          & PL + absorption                  & $1.50$ (fix)               & $0.08\sp{+0.05}\sb{-0.04}$ &                            &      & 7.7   &       \\[1.0ex]
          & PL + partially covering absorber & $1.50$ (fix)               & $4.96\sp{+11.2}\sb{-3.46}$ & $0.23\sp{+0.07}\sb{-0.08}$ &      & 18.4  & 99.7  \\[1.0ex]
          & PL + ionized absorption          & $1.50$ (fix)               & $0.58\sp{+1.03}\sb{-0.41}$ &      & $2.09\sp{+0.48}\sb{-0.71}$ & 12.7  & 97.6  \\[2.5ex]

 \chandra & Power law                        & $1.01\sp{+0.06}\sb{-0.05}$ &                            &                            &      & 8.3   &       \\[1.0ex]
 2005     & PL + absorption                  & $1.20\sp{+0.11}\sb{-0.10}$ & $0.19\sp{+0.10}\sb{-0.08}$ &                            &      & 36.5  & 99.9  \\[1.0ex]
          & PL (2--8 keV)                    & $1.42\sp{+0.20}\sb{-0.20}$ &                            &                            &      & 19.8  &       \\[1.0ex]
          & PL + absorption                  & $1.42$ (fix)               & $0.35\sp{+0.07}\sb{-0.06}$ &                            &      & 12.8  &       \\[1.0ex]
          & PL + partially covering absorber & $1.42$ (fix)               & $1.40\sp{+0.84}\sb{-0.59}$ & $0.60\sp{+0.10}\sb{-0.07}$ &      & 36.9  & 99.8  \\[1.0ex]
          & PL + ionized absorption          & $1.42$ (fix)               & $0.55\sp{+0.51}\sb{-0.13}$ &      & $0.89\sp{+0.58}\sb{-0.31}$ & 18.7  & 94.4  \\[2.5ex]

 \xmm     & Power law                        & $1.19\sp{+0.03}\sb{-0.03}$ &                            &                            &      & 27.6  &       \\[1.0ex]
 2013     & PL + absorption                  & $1.27\sp{+0.05}\sb{-0.05}$ & $0.08\sp{+0.05}\sb{-0.05}$ &                            &      & 38.7  & 99.5  \\[1.0ex]
          & PL (2--10 keV)                   & $1.33\sp{+0.09}\sb{-0.09}$ &                            &                            &      & 44.5  &       \\[1.0ex]
          & PL + absorption                  & $1.33$ (fix)               & $0.12\sp{+0.03}\sb{-0.03}$ &                            &      & 34.2  &       \\[1.0ex]
          & PL + partially covering absorber & $1.33$ (fix)               & $0.92\sp{+1.28}\sb{-0.66}$ & $0.31\sp{+0.27}\sb{-0.08}$ &      & 40.0  & 96.3  \\[1.0ex]
          & PL + ionized absorption          & $1.33$ (fix)               & $0.26\sp{+0.42}\sb{-0.21}$ &      & $1.42\sp{+0.66}\sb{-2.94}$ & 36.4  & 87.2  \\[2.5ex]

 \chandra & Power law                        & $1.02\sp{+0.07}\sb{-0.07}$ &                            &                            &      & 45.1  &       \\[1.0ex]
 2014     & PL + absorption                  & $1.16\sp{+0.12}\sb{-0.12}$ & $0.17\sp{+0.13}\sb{-0.12}$ &                            &      & 63.3  & 98.9  \\[1.0ex]
          & PL (2--8 keV)                    & $1.38\sp{+0.21}\sb{-0.21}$ &                            &                            &      & 93.1  &       \\[1.0ex]
          & PL + absorption                  & $1.38$ (fix)               & $0.36\sp{+0.10}\sb{-0.09}$ &                            &      & 36.4  &       \\[1.0ex]
          & PL + partially covering absorber & $1.38$ (fix)               & $1.57\sp{+1.23}\sb{-0.77}$ & $0.54\sp{+0.14}\sb{-0.09}$ &      & 62.2  & 99.7  \\[1.0ex]
          & PL + ionized absorption          & $1.38$ (fix)               & $0.71\sp{+0.88}\sb{-0.30}$ &      & $1.27\sp{+0.55}\sb{-0.64}$ & 51.2  & 97.8  \\[2.5ex]
     \hline\\
    \end{tabular}
    \end{center}
    \vspace*{-0.5cm}
    \begin{raggedright}
    \hspace*{1.95cm} \hangindent=2cm \rightskip=2cm \textbf{Notes.}  The spectral fits are carried out using data in the 
    energy range $0.5-10$~keV for \xmm and $0.5-8$~keV for \textit{Chandra}, unless stated otherwise.  Parameters are 
    quoted with 90\% errors.  $\xi=L/nr\sp{2}$ with units of ${\rm erg\,s}\sp{-1}\,{\rm cm}$; \logxi$=-3$ for neutral 
    material, and \logxi$=+6$ for fully ionized material.  H0 values greater than 1\% indicate the model is an acceptable fit 
    to the data.  $F$-test values greater than 99\% indicate that the model including the additional parameter is 
    statistically preferred.  Despite differences in the fitting methodology, the best-fitting parameters for the 
    partially covering absorber model for the 2005 \chandra data are consistent with those given by M06.\\
    \end{raggedright}
    \vspace*{0.4cm}
\end{table*}

The \mbox{X-ray} detectors of ACIS and EPIC are known to yield different spectral fitting results even
during simultaneous observations of the same object, or in observations of non-variable objects
(e.g.,~galaxy clusters; \citealt{nevalainen10}).  By comparison with the typical \gmm and \nh differences
found in simultaneous ACIS and EPIC observations of the BL~Lac PKS~2155-304 \citep{ishida11}, we conclude
that the 2013 \xmm results are likely to be consistent with both the 2014 and 2005 \chandra observations 
after accounting for these cross-calibration differences.  Only the original 2003 \xmm observation appears
to show \pgten\/ with different spectral parameters ($\Gamma$, $N\sb{\rm H}$).  This was previously observed
and discussed in M06.

However, the detectability of absorption is dependent on the spectral quality.  As the 2003 \xmm spectra 
contain fewer counts than are required to significantly detect absorption of $N\sb{\rm H}\sim0.08$\nhunits\, 
(the level measured in the 2013 \xmm data), 100\% of the time (see \citealt{scott12} for detectability 
curves), this may be limiting our ability to significantly detect absorption even though it is present 
in the source.  If instead, we fix \gmm to the value obtained from the fit to high energies ($\ge 2$~keV), 
as shown in Figure~\ref{fig:xray_spectra}, we do measure a significant absorption component with 
$N\sb{\rm H,\,2003}=(0.08\sp{+0.05}\sb{-0.04})$\nhunits.  This is consistent with the absorption measured 
in the 2013 \xmm data using the same spectral modeling.  We therefore suggest that the spectral properties 
of \pgten\/ indicated in both \xmm observations may be consistent, and therefore the amount of intrinsic 
absorption has not varied dramatically between all 4 observations.

If each of the data sets are modeled with an absorbed power-law in which \gmm is fixed to the values obtained
from the fit to high energies ($\ge 2$~keV), steeper power-law slopes and higher absorption columns are 
obtained (see Table~\ref{table:fits}).  However, we note that the \gmm values remain slightly flatter than 
expected (e.g.,~\citealt{reeves00}), which may indicate that the shape of the underlying continuum above 
$2$~keV is still affected by the absorption.  If the spectra are modeled with a power-law using only energies 
greater than $4$~keV, slightly steeper spectral indices are obtained but they remain formally consistent 
($\lesssim1\sigma$) with the previous estimates.  Adopting these steeper \gmm values in the subsequent 
absorption modeling leads to a possible increase in the absorption column density of a factor $\lesssim2$.

Thus far the spectral modeling has only considered absorption due to neutral material that is fully 
covering the central source.  While this is useful for a simple and consistent quantification of 
the amount of absorption present in each observation, the presence of more complex \mbox{X-ray} absorption 
is thought to be common in both RQ and RL BAL quasars (e.g.,~\citealt{gallagher02,gallagher06,brotherton05}).  
We therefore also fit the spectra with a partially covering absorption model (\texttt{tbabs*zpcfabs*po} 
in \textsc{xspec}) and a partially ionized absorption model (\texttt{tbabs*zxipcf*po}), in which 
\textsc{zxipcf} \citep{reeves08} is implemented with $f$ fixed at 1.0 to model a partially ionized, 
but fully covering, absorber.  We fix the \gmm value to that found for high energies to ensure constrained 
$N\sb{\rm H}$, $\xi$ and $f$ parameters.  These are listed in Table~\ref{table:fits}, along with 
null-hypothesis probabilities (H0), which if greater than 1\% indicate the model is an acceptable 
fit to the data, and $F$-test probabilities, for which we say values greater than 99\% indicate 
that the model including the additional component is statistically preferred.

In all 4 observations, the $F$-test probability and H0 values are higher for the partially covering 
absorption model than the partially ionized absorption model, although both provide acceptable fits to 
the data.  For both models the best-fitting spectral parameters are consistent when comparing the 
2 \chandra observations or the 2 \xmm observations.  This again suggests that the properties of 
the \mbox{X-ray} absorption have not changed significantly on long timescales ($\sim 8$~yrs in the source 
rest-frame).

We test for the presence of iron emission lines in each of our data sets by adding Gaussians to a 
power-law fit at high energies, i.e.,~the model \texttt{tbabs*(po+zgauss)}.  We use the Cash statistic 
and unbinned spectra as the increased bin size in the grouped data could limit the detection of a narrow 
line.  We fix the width of the line to $\sigma=0.01$~keV and fix the rest-frame energy of the line to the 
values $6.4$ (for neutral iron), $6.7$ (He-like), and $6.9$~keV (H-like).  In agreement with M06 and 
\citet{luo13}, we find no strong evidence for iron emission lines at these energies.  We obtain only upper 
limits on the EW in all cases except for the $6.4$~keV line in the 2013 \xmm spectra.  This is unsurprising 
as these spectra have the highest spectral resolution and highest number of counts of all those we consider.
The EW measured is ${\rm EW\sb{\rm 6.4~keV}}=150\sp{+80}\sb{-60}$~eV, and the line can be seen in the
residuals shown in Figure~\ref{fig:xray_spectra}.  We also consider a joint fitting of all 6
\mbox{X-ray} spectra (MOS, pn, and ACIS) and obtain ${\rm EW\sb{6.4~keV}=90\pm40~{\rm eV}}$ (consistent
with the upper limit of ${\rm EW\sb{6.4~keV}<105~{\rm eV}}$ obtained by M06),
${\rm EW\sb{6.7~keV}<70~{\rm eV}}$, and ${\rm EW\sb{6.9~keV}<70~{\rm eV}}$,
suggesting that a low level of neutral iron emission is present.\\


\subsection{Broad Absorption Line Variability}
\label{section:uv}
In this section we compare our newly obtained NUV spectrum from \hst COS to archival spectra taken by 
\iue in 1982 and 1986, and a \hst Space Telescope Imaging Spectrograph (STIS; \citealt{STIS}) spectrum 
taken in 2003, in order to look for variability of the \civ\/ BAL.  Details of all 4 observations are 
given in Table~\ref{table:uv_bal}.  We also include the S/N of the spectra and the observed UV continuum 
fluxes, measured between $\lambda\sb{\rm obs}=1750-1825$~\AA\/ ($\lambda\sb{\rm rest}=1411-1471$~\AA).  
With the exception of a $\sim2\sigma$ flux decrease between the 2003 and 2014 \hst spectra, there does 
not appear to be significant variation in the UV flux.

\begin{table*}[ht!]
\centering
\caption{UV Observations and BAL Parameters}
\label{table:uv_bal}
\vspace*{-0.4cm}
    \begin{center}
    \begin{tabular}{lcccc}
      \hline \hline\\[-2.0ex]
      \multicolumn{1}{c}{ } &
      \multicolumn{1}{c}{Observation 1} &
      \multicolumn{1}{c}{Observation 2} &
      \multicolumn{1}{c}{Observation 3} &
      \multicolumn{1}{c}{Observation 4}\\[0.5ex]
      \hline\\[-1.0ex]
          Date                                        &     1982-04-24 &      1986-01-12 &     2003-03-31 &      2014-01-20 \\[1.0ex]
          MJD                                         &          45083 &           46442 &          52729 &           56678 \\[1.0ex]
          Mission                                     &           \iue &            \iue &           \hst &            \hst \\[1.0ex]
          Instrument                                  &         \ldots &          \ldots &           STIS &             COS \\[1.0ex]
          Grating                                     &         \ldots &          \ldots &          G230L &           G230L \\[1.0ex]
          Central wavelength (\AA)                    &         \ldots &          \ldots &           2376 &            2950 \\[1.0ex]
          Resolution (\AA)                            &        $\sim6$ &         $\sim6$ &        $\sim5$ &         $\sim1$ \\[1.0ex]
          Exposure time (s)                           &          22260 &           18900 &           2292 &            2562 \\[1.0ex] 
          UV flux ($\times10\sp{-15}\,{\rm erg\:s}\sp{-1}\,{\rm cm}\sp{-2}$)               
                                                      &  $9.29\pm0.87$ &   $8.03\pm1.01$ &  $8.53\pm0.41$ &   $7.20\pm0.49$ \\[1.0ex]
          Signal-to-noise (S/N)                       &         $10.6$ &           $7.7$ &         $20.6$ &          $14.6$ \\[1.0ex]
          Mini-BAL EW (\AA)                           &  $3.38\pm0.32$ &   $4.14\pm0.38$ &  $2.74\pm0.25$ &   $3.38\pm0.42$ \\[1.0ex]
          BAL EW (\AA)                                &  $3.03\pm0.56$ &  $13.38\pm0.64$ &  $2.98\pm0.34$ &  $11.24\pm0.56$ \\[1.0ex]
          {\it Comparison of BAL to 2014:}            &                &                 &                &                 \\[1.0ex]
          \hspace{1cm}$\tau$ (days)                   &           9346 &            8251 &           3183 &          \ldots \\[1.0ex]
          \hspace{1cm}$\Delta EW$ (\AA)               &  $8.21\pm0.79$ &  $-2.14\pm0.85$ &  $8.26\pm0.66$ &          \ldots \\[1.0ex]
          \hspace{1cm}$\langle EW \rangle$ (\AA)      &  $7.14\pm0.40$ &  $12.31\pm0.43$ &  $7.11\pm0.34$ &          \ldots \\[1.0ex]
          \hspace{1cm}$\Delta EW/ \langle EW\rangle$  &  $1.15\pm0.13$ &  $-0.17\pm0.07$ &  $1.16\pm0.11$ &          \ldots \\[1.0ex]
          {\it Comparison of BAL to 2003:}            &                &                 &                &                 \\[1.0ex]
          \hspace{1cm}$\tau$ (days)                   &           6163 &            5068 &         \ldots &          \ldots \\[1.0ex]
          \hspace{1cm}$\Delta EW$ (\AA)               & $-0.05\pm0.66$ & $-10.40\pm0.72$ &         \ldots &          \ldots \\[1.0ex]
          \hspace{1cm}$\langle EW \rangle$ (\AA)      &  $3.01\pm0.33$ &   $8.18\pm0.36$ &         \ldots &          \ldots \\[1.0ex]
          \hspace{1cm}$\Delta EW/ \langle EW\rangle$  & $-0.02\pm0.26$ &  $-1.27\pm0.10$ &         \ldots &          \ldots \\[1.0ex]
          {\it Comparison of BAL to 1986:}            &                &                 &                &                 \\[1.0ex]
          \hspace{1cm}$\tau$ (days)                   &           1095 &          \ldots &         \ldots &          \ldots \\[1.0ex]
          \hspace{1cm}$\Delta EW$ (\AA)               & $10.35\pm0.85$ &          \ldots &         \ldots &          \ldots \\[1.0ex]
          \hspace{1cm}$\langle EW \rangle$ (\AA)      &  $8.21\pm0.43$ &          \ldots &         \ldots &          \ldots \\[1.0ex]
          \hspace{1cm}$\Delta EW/ \langle EW\rangle$  &  $1.26\pm0.12$ &          \ldots &         \ldots &          \ldots \\[1.5ex]
     \hline
    \end{tabular}
    \end{center}
    \vspace*{-0.2cm}
    \begin{raggedright}
    \hspace*{3.15cm} \hangindent=3.3cm \rightskip=3.3cm \textbf{Notes.} The UV flux listed is the observed flux between 
    $\lambda\sb{\rm obs}=1750-1825$~\AA\/ ($\lambda\sb{\rm rest}=1411-1471$~\AA) located bluewards of the \ion{C}{4} BAL.  
    The bottom rows compare the BAL properties measured in each of the spectra.  
    $\tau$ is the time between observations in the rest-frame of \pgten\, and is calculated as 
    $\tau=({\rm MJD}\sb{\rm new}-{\rm MJD}\sb{\rm old})/(1+z)$.  $\Delta {\rm EW}={\rm EW}\sb{\rm new}-{\rm EW}\sb{\rm old}$ 
    is the change in the EW, and $\langle EW\rangle=0.5(EW\sb{\rm new}+EW\sb{\rm old})$ is the average EW.  The ratio of 
    these two quantities gives the fractional change listed.\\
    \end{raggedright}
    \vspace*{0.4cm}
\end{table*}

\begin{figure}[h]
  \centering 
  \vspace*{0.2cm}
    \includegraphics[width=0.48\textwidth]{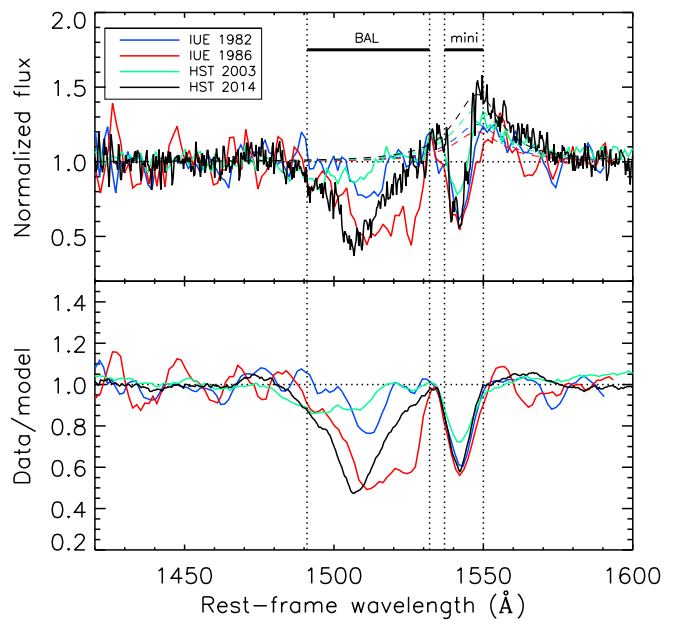}
    \vspace*{0.05cm}
  \caption{UV spectra of \pgten, covering the \civ\/
    emission/absorption-line region.  Archival \iue spectra from 1982
    and 1986 are shown in blue and red respectively, and an archival
    \hst STIS spectrum from 2003 is shown in green.  Our recent \hst
    COS spectrum (2014 Jan) is shown in black.  The top panel shows
    each of the spectra normalized to the same scale.  The dashed
    lines show the best-fitting model for each spectrum, composed of a
    linear component to model the local continuum, and a Voigt profile
    for the \civ\/ emission line.  The bottom panel shows the
    resulting ratio spectra after each spectrum has been divided by
    its best-fitting model.  They have been smoothed with a Gaussian
    of 8~\AA\/ width.  The regions over which the EW of the BAL and
    mini-BAL are determined are shown by the vertical, black, dotted
    lines.\\}
  \label{fig:bal}
\end{figure}

Each of the 4 spectra were modeled with a Voigt profile and a linear function to represent the \civ\/ 
emission line and the local continuum, respectively.  These models are shown by the dashed lines in 
the top panel of Figure~\ref{fig:bal}.  The spectra were then divided by the best-fitting model, 
resulting in the ratio spectra shown in the bottom panel of Figure~\ref{fig:bal}.  

This shows a mini-BAL with a FWHM of $1553~\kms$ centered on $\lambda\sb{\rm rest}\sim1542$~\AA.  
The offset of the absorption from the expected emission wavelength of \civ\/ ($\lambda=1551$~\AA) 
corresponds to a typical outflow velocity of \mbox{$v\sb{\rm out}\sim1745~\kms$ for the mini-BAL}.  
The BAL has a FWHM of $3372~\kms$ (in the 2014 spectrum), indicating a large range of velocities 
observed along the line-of-sight within the accelerated outflow.  These range from 
\mbox{$v\sb{\rm out}\sim3700~\kms$} to \mbox{$v\sb{\rm out}\sim11800~\kms~(\sim0.04c)$}, with the 
velocity of the deepest part of the BAL trough being at $v\sb{\rm out}\sim8000~\kms$.

The EW of the BAL and mini-BAL are directly measured from the ratio 
spectra.  The edges of the BAL region are defined as the wavelengths at which the BAL reaches a 
level which is 90\% that of the continuum following the standard definition of \citet{weymann91}.  
In the 2014 spectrum this corresponds to a range of $\lambda\sb{\rm rest}=1491-1532$~\AA\/ for the BAL, and 
$\lambda\sb{\rm rest}=1537-1550$~\AA\/ for the mini-BAL.  The EW is determined simply as a sum of the 
flux in all pixels within this region, and the error estimate includes contributions from the continuum 
placement error, statistical spectral noise, and the uncertainty in the region used.  The EW of the BAL 
and mini-BAL in the newly obtained 2014 spectrum are $11.24\pm0.56$~\AA, and $3.38\pm0.42$~\AA, respectively.  
These are listed in Table~\ref{table:uv_bal} along with EW estimates for the 3 archival spectra determined 
by the same method\footnote{We use a different normalization and modeling of the spectra to that of 
\citet{welling14}, resulting in slightly different EW values.}.  The final rows of Table~\ref{table:uv_bal} 
compare the EW values of the BAL measured in each of the spectra, with the fractional change in EW being given by 

\begin{equation}
\frac{\Delta EW}{\langle EW\rangle} = \frac{EW\sb{\rm new} - EW\sb{\rm old}}{0.5(EW\sb{\rm new} + EW\sb{\rm old})}.
\vspace*{0.25cm}
\end{equation} 

Figure~\ref{fig:bal} (bottom) shows the \civ\/ BALs for each of the 4 spectra, smoothed with a Gaussian 
of 8~\AA\/ width.  This clearly shows the strongly variable \civ\/ absorption that \pgten\/ 
is known to possess, i.e.,~the BAL strength increases from the 1982 \iue observation (blue) to the 1986 
\iue observation (red), before diminishing greatly in depth in the 2003 \hst observation (green).  
Further, dramatic variability is revealed by the latest 2014 \hst observation (black), which shows the BAL 
has deepened.  In addition to large changes in EW, the shape of the BAL trough also varies significantly 
between the observations.  In comparison, the mini-BAL at a smaller blueshift shows little variability 
in both strength or shape, with the exception of the 2003 spectrum in which the trough is much shallower.

The 2014 \hst spectrum does not cover the wavelength of other possible BAL troughs (e.g.,~\ovi, \nv),
nor do we see evidence for a \siiv\/ BAL.\\

\section{Extra-nuclear X-ray Emission}
\label{section:image}
The 2005 \chandra observation of \pgten\/ revealed both \mbox{X-ray} emission $\sim8$\arcsec\/ SW of
the nucleus, thought to be related to the FRI radio jet, and diffuse \mbox{X-ray} emission surrounding
the source and extending out to $\sim 40-50$\arcsec\/ (M06).  In this work, we combine the data from
both \chandra observations giving an increased exposure time of $\sim 100$~ks ($2.5\times$ deeper than
previously) with which to further investigate these features.

\subsection{X-ray Jet Emission}
\label{section:jet}
\begin{figure*}[ht!]
  \centering
  \begin{tabular}{cc}
    \includegraphics[width=0.48\textwidth]{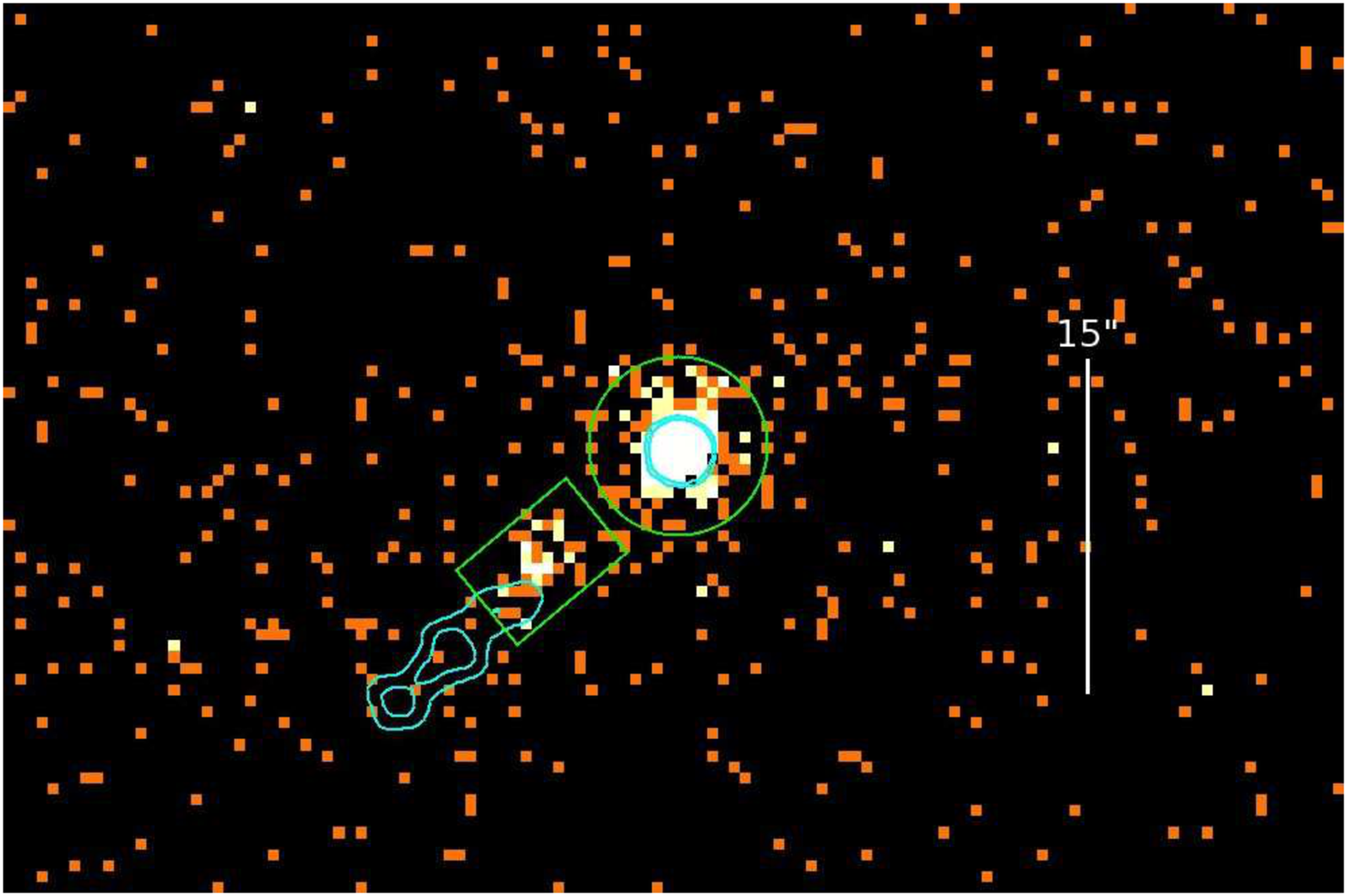} &
    \includegraphics[width=0.48\textwidth]{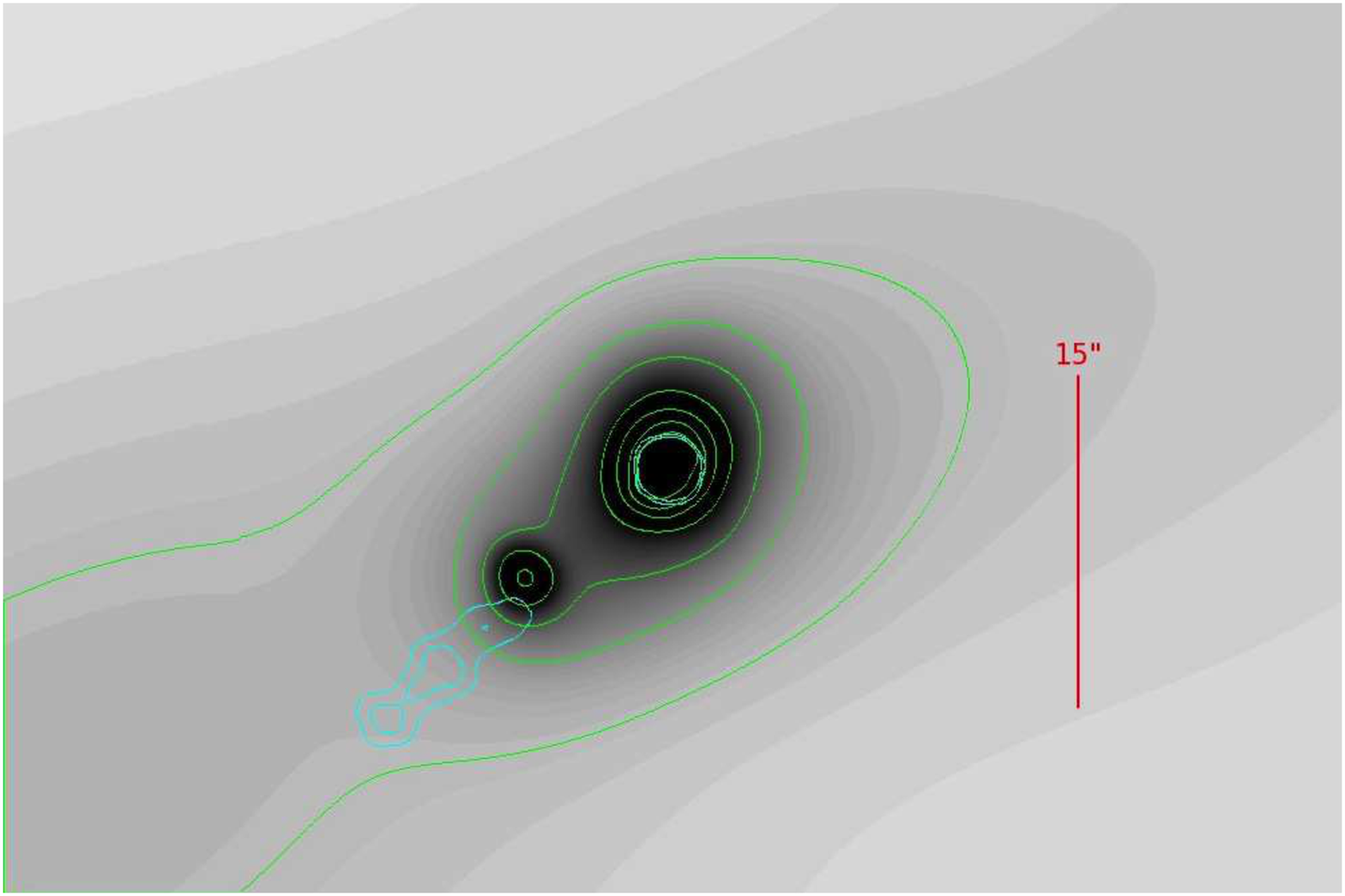}\\
    \includegraphics[width=0.48\textwidth]{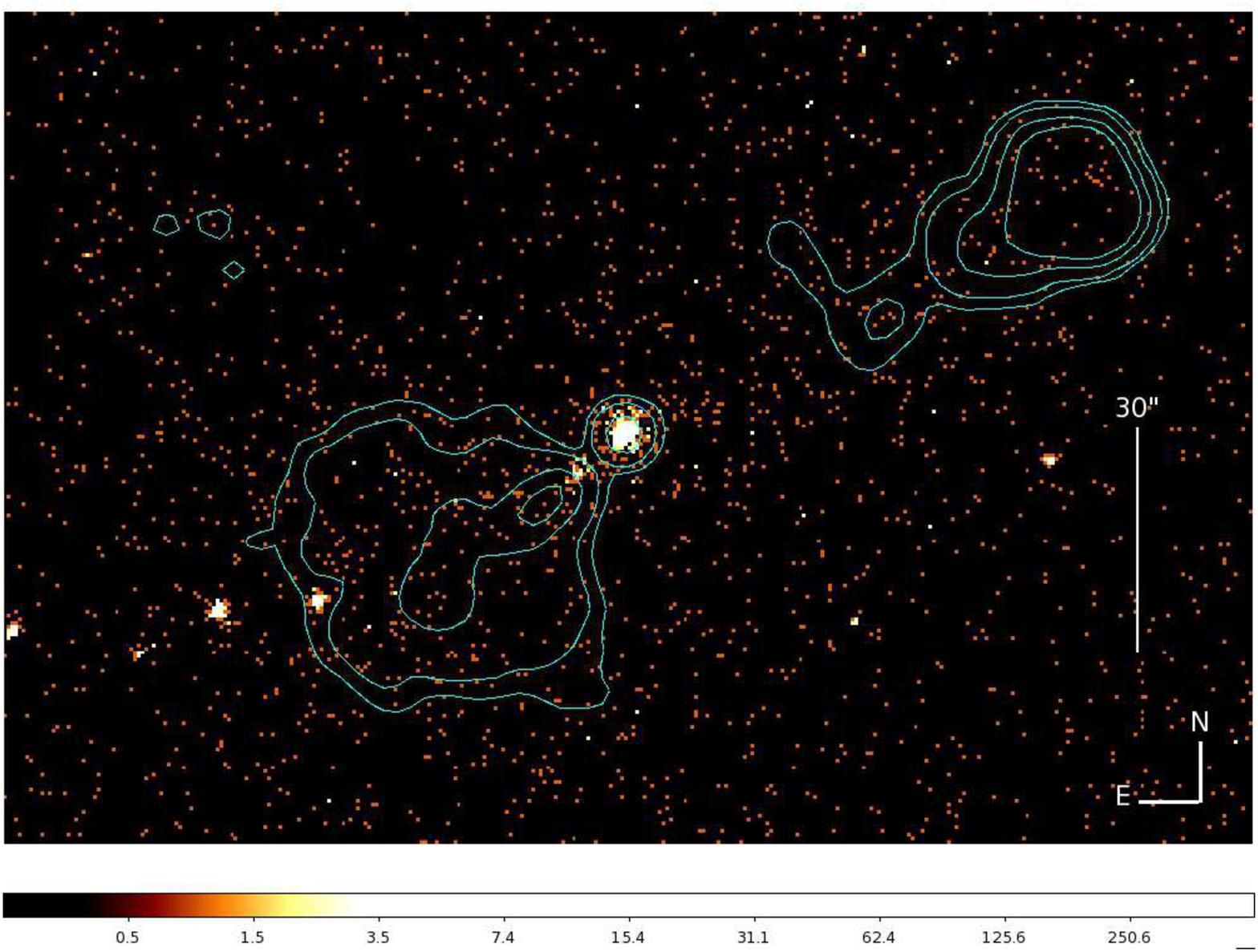} &
    \includegraphics[width=0.48\textwidth]{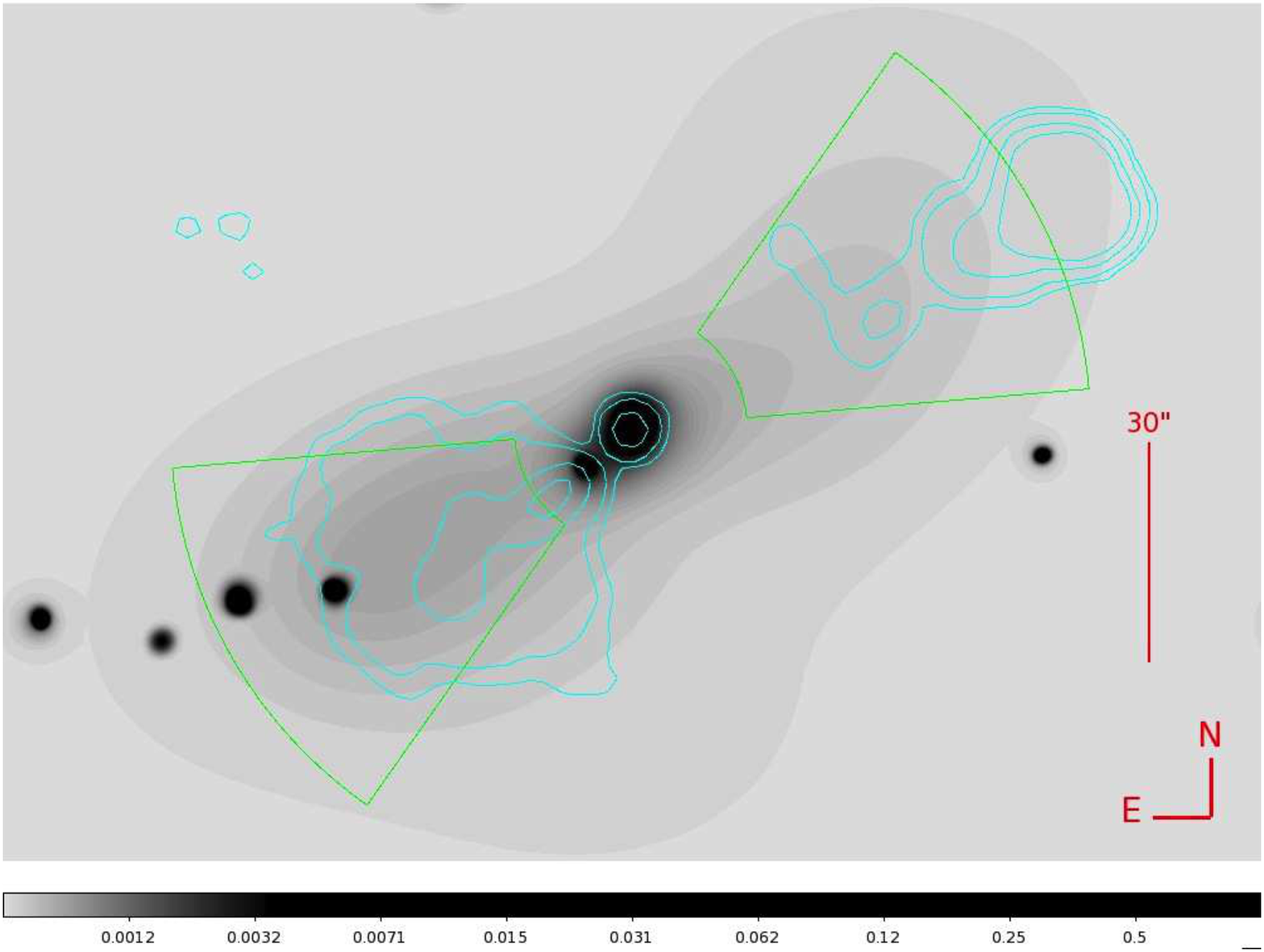}\\
    \vspace*{0.1cm}
  \end{tabular}
  \caption{Top left -- $0.5-4$~keV \mbox{X-ray} image from the
    combined 2005 and 2014 \chandra observations.  The \mbox{X-ray}
    emitting nuclear source and the \mbox{X-ray} knot in the SE jet
    are clearly visible, and the extraction regions used for the
    nuclear source (circle) and the jet (rectangle) are shown in
    green.  Radio contours from the VLA at 4.9~GHz (6~cm)
    corresponding to 0.75 and 1.25~mJy beam$\sp{-1}$ are overlaid in
    blue.  Top right -- An adaptively smoothed version of the top left
    image.  Shown in green are \mbox{X-ray} contours corresponding to
    count rates of 0.0005, 0.001, 0.002, 0.005, 0.01, 0.02, and 0.05
    counts per square pixel.  Blue contours show the same VLA data as
    in the top left image.  Bottom left -- The same $0.5-4$~keV
    \mbox{X-ray} image shown in the top left image, but including the
    extended regions of the system.  Blue contours show lower angular
    resolution radio data from FIRST (at 1.4~GHz/21~cm) at 1.5, 3, 8,
    and 15~mJy beam$\sp{-1}$.  Bottom right -- An adaptively smoothed
    version of the bottom left image.  The black dots to the SE and W
    are unrelated \mbox{X-ray} sources, likely background AGN.  The
    green wedges correspond to those used to determine the amount of
    diffuse \mbox{X-ray} emission along the paths of the SE jet and NW
    counterjet.  The blue contours show the same FIRST data as in the
    bottom left image.  At the redshift of \pgten, 15\arcsec\/
    corresponds to a projected physical distance of 57~kpc.\\}
  \label{fig:images}
\end{figure*}

Figure~\ref{fig:images} (top left) shows the $0.5-4$~keV\footnote{There are few counts at energies
greater than 4~keV.} \mbox{X-ray} image produced from the addition of the 2005 and 2014 \chandra
observations.  The 2014 data were reprojected onto the co-ordinate system of the 2005 data to account
for a 0.7\arcsec\/ offset between the source positions.  The spectral extraction regions used for the
nuclear source and the \mbox{X-ray} jet emission are indicated by the green circle and rectangle,
respectively.  The blue contours are produced from 4.9~GHz (6~cm) VLA radio data \citep{fomalont81}
with a resolution of 1.5\arcsec, and show flux density levels of 0.75 and 1.25~mJy beam$\sp{-1}$.
Figure~\ref{fig:images} (bottom left) shows the same \mbox{X-ray} image, but on a larger scale.  In
this case, radio data from the FIRST survey \citep{becker95} at 1.4~GHz (21~cm) with a resolution 
of 5.4\arcsec\/ are used to give contours at 1.5, 3, 8, and 15~mJy beam$\sp{-1}$ (overlaid in blue).
This clearly shows the HYMOR \citep{gopal00} nature of the radio emission, with the SE jet showing a
decreasing surface-brightness (FRI), whereas the NW counterjet shows an increase in surface brightness
at larger radii from the nucleus, typical of the FRII class.  \mbox{X-ray} emission is observed
following the same direction as the SE radio jet, albeit at closer radial distances than the peak of
the radio emission.  

Figure~\ref{fig:images} (top right) shows the central region of the combined $0.5-4$~keV \mbox{X-ray}
image which has been rebinned to a subpixel size of $0.05\arcsec$ and adaptively smoothed using a
Gaussian kernel to achieve a minimum S/N of 3.  \mbox{X-ray} contours corresponding to 0.0005, 0.001,
0.002, 0.005, 0.01, 0.02, and 0.05 counts per square pixel are overlaid in green.  These correspond to
flux densities between $6.6\times10\sp{-35}$ and
$6.6\times10\sp{-33}\,{\rm erg\,s}\sp{-1}\,{\rm cm}\sp{-2}\,{\rm Hz}\sp{-1}\,{\rm arcsec}\sp{-2}$,
assuming a photon energy of 1~keV, for which the ACIS QE is $\sim0.5$ and the effective area is 
$\sim400~\textrm{cm}\sp{2}$.\footnote{\url{http://cxc.harvard.edu/proposer/POG/html/chap6.html}
  (\chandra handbook, chapter 6)}  This image again clearly shows the \mbox{X-ray} knot which lies
upstream of the radio jet emission at a distance of $\sim 8\arcsec$ (projected distance of 30~kpc) 
and position angle of 130\degree\/ from the nuclear source.  No similar \mbox{X-ray} emission is 
observed in relation to the NW counterjet.

Spectra of the \mbox{X-ray} jet emission to the SE of the nuclear source are extracted from each
individual \chandra observation using the rectangular region shown on Figure~\ref{fig:images} (top
left).  There are 29 \mbox{X-ray} counts ($0.5-8$~keV) in both the 2005 and 2014 jet spectra, $\sim2-3$
of which are likely background counts, with some due to the larger-scale diffuse \mbox{X-ray}
emission (see Section~\ref{section:diffuse}).  The spectra are modeled with a power law absorbed only by 
a Galactic contribution.  The 2005 spectrum is relatively flat with $\Gamma=1.63\sp{+0.75}\sb{-0.57}$,
whereas the 2014 spectrum is comparatively steep, $\Gamma=2.21\sp{+0.99}\sb{-0.79}$.  However, due to
the large (90\%) errors, this variation is not significant ($0.9\sigma$).  We also simultaneously model
both spectra to obtain a better constraint on the power-law slope; $\Gamma=1.96\sp{+0.63}\sb{-0.52}$.
A $1.6\sigma$ decrease in the flux of the jet is measured, falling from  
$F\sb{\rm 0.5-8\,keV,\,2005}=(5.47\sp{+2.09}\sb{-1.55})\times10\sp{-15}\flux$ to 
$F\sb{\rm 0.5-8\,keV,\,2014}=(2.86\sp{+2.04}\sb{-1.12})\times10\sp{-15}\flux$.  
This variation is largely due to the difference in the best-fitting model as when \gmm is fixed to 1.96
for each observation, the measured fluxes differ by only $\sim1\sigma$ and show a decrease of $\sim30$\%.  

\begin{figure*}[ht!]
  \centering
    \begin{tabular}{cc}
      \includegraphics[width=0.48\textwidth]{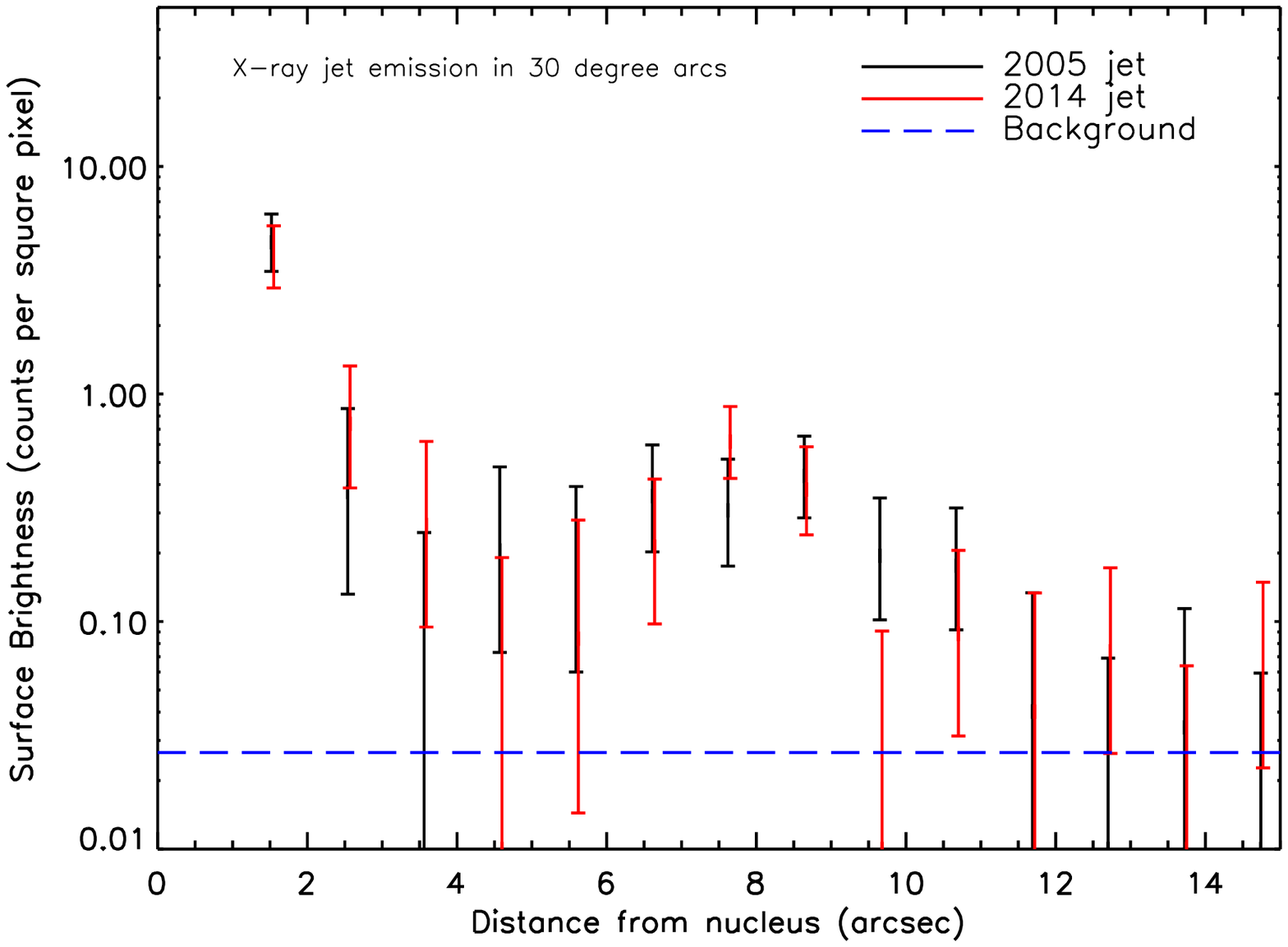} &
      \includegraphics[width=0.48\textwidth]{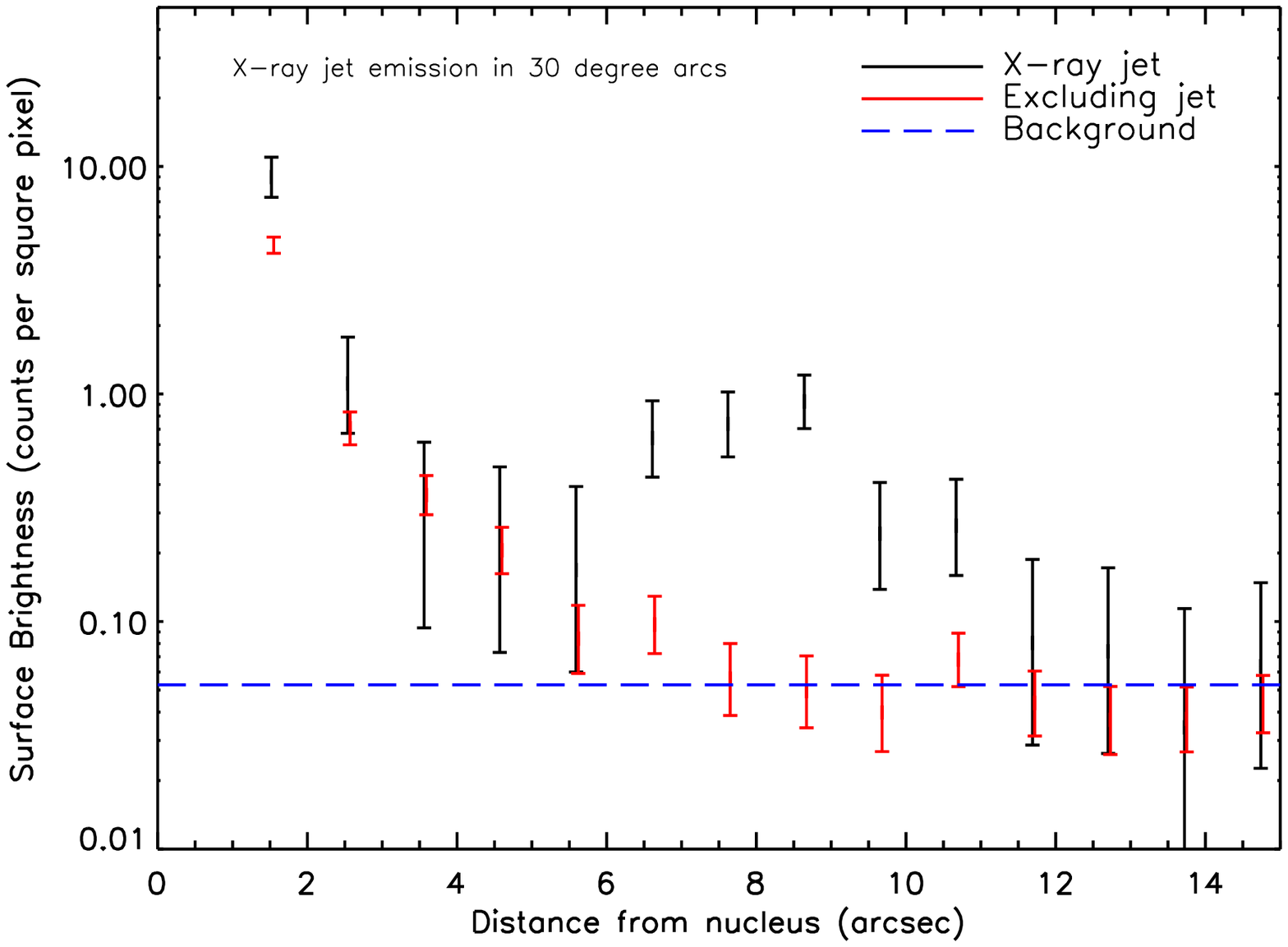} \\
      \vspace*{0.07cm}
    \end{tabular}
      \hspace*{1cm}
      \includegraphics[width=0.8\textwidth]{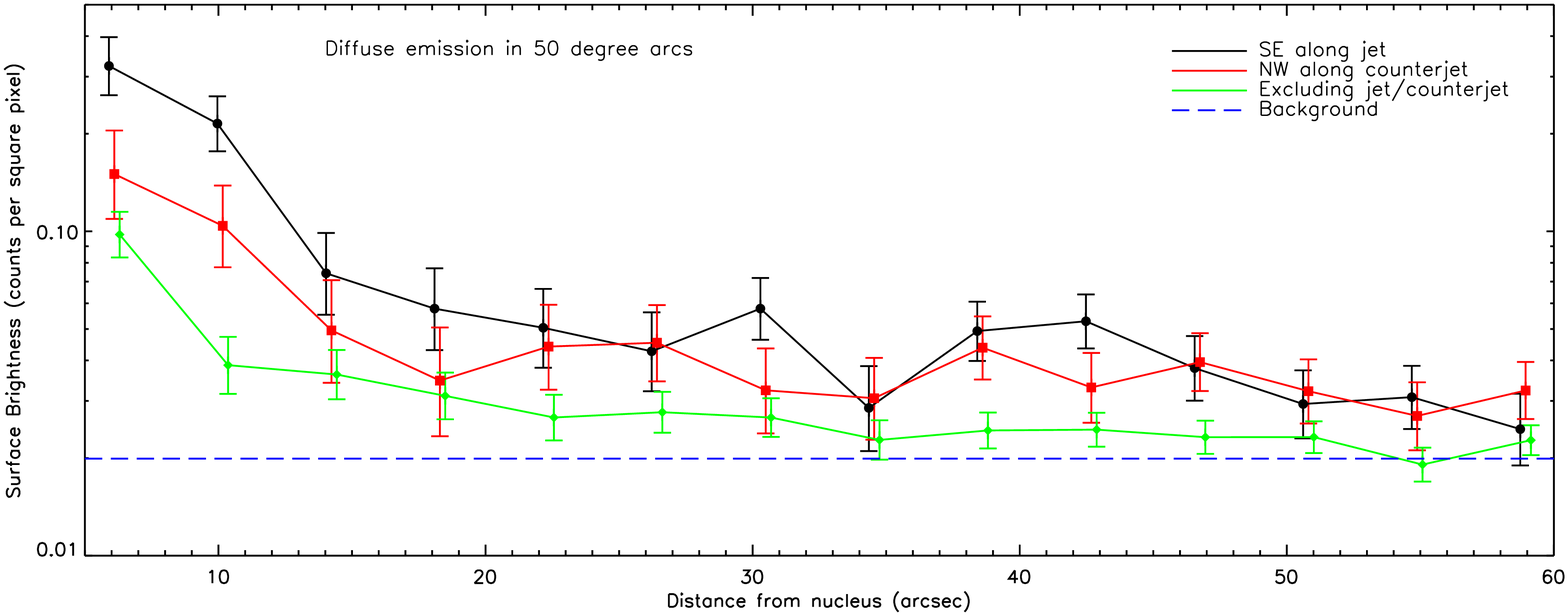}\\
      \vspace*{0.1cm}
  \caption{Radial surface brightness profiles.  In each case the
    profiles are extracted from the $0.5-4$~keV \chandra images which
    were rebinned to a sub-pixel bin size of $0.05$\arcsec.  Surface
    brightnesses are calculated as the number of counts per square
    pixel and the \citet{gehrels86} formulation of errors is used.
    Top left -- A comparison of the SE jet profiles for the 2005 data
    (black) and the 2014 data (red), extracted from concentric
    30\degree\/ arcs at angles of $115-145$\degree, centered on the
    source position, and with a thickness of 1\arcsec.  The blue
    dashed line gives an indication of the background level ($\sim
    0.02$ counts per square pixel in 2005, $\sim 0.03$ counts per
    square pixel in 2014) which is determined from an
    $8\arcsec\times8\arcsec$ box.  Top right -- The SE jet profile
    extracted from the combined 2005 and 2014 \chandra observations
    (black), using the same regions as in the top left panel.  The
    profile shown in red is extracted from the remaining 330\degree\/
    annular sectors for comparison.  The blue dashed line indicates
    the background level ($\sim 0.05$ counts per square pixel in the
    combined data).  Bottom -- Profiles of the diffuse \mbox{X-ray} emission
    extracted from concentric 50\degree\/ arcs to the SE (black)
    and NW (red).  These are compared to a profile extracted from the
    remaining 260\degree\/ from each concentric ring (green).  Fifteen
    equal-thickness rings are used from $0-60$\arcsec\/ radial
    distance from the central source position.  The blue, dashed line
    indicates the background level (0.02 counts per square pixel)
    determined from a $30\arcsec\times30\arcsec$ box.\\}
  \label{fig:radial}
\end{figure*}

Radial surface-brightness profiles for the SE \mbox{X-ray} jet emission were created from the $0.5-4$~keV
\mbox{X-ray} images which were first rebinned to a sub-pixel size of $0.05\arcsec$.  Concentric annular
sectors covering a 30\degree\/ arc from $115-145$\degree, with a thickness of $1\arcsec$ were used to
determine the jet profile.  Figure~\ref{fig:radial} (top left) compares the jet profiles from the 2005
(black) and 2014 (red) observations.   This shows that the distribution of the \mbox{X-ray} emission has
not varied significantly between the 2 epochs.  An increase in the \mbox{X-ray} surface brightness is
clear at $\sim 8\arcsec$ from the nuclear source, corresponding to the peak of the \mbox{X-ray} jet
emission.  This knot is also clearly extended by $\sim 4\arcsec$.  Figure~\ref{fig:radial} (top right)
shows the jet profile from the combined 2005 and 2014 data (black).  This is compared to a profile created
from the remaining 330\degree\/ of the annular sectors (red).  This clearly shows the excess \mbox{X-ray}
surface brightness to the SE, which is significantly higher than the background level (estimated from a
$8\arcsec\times8\arcsec$ box and indicated by the dashed blue line).\\  

\subsection{Diffuse X-ray Emission}
\label{section:diffuse}
Figure~\ref{fig:images} (bottom right) shows the wider-scale \mbox{X-ray} image, smoothed in the
manner described in the previous section.  This image shows more clearly the low-level diffuse
\mbox{X-ray} emission surrounding \pgten\/ which was previously reported by M06.  The emission
largely appears to follow the direction of the SE FRI jet and the NW FRII counterjet, rather than
forming a spherical halo around the nuclear source.  The diffuse \mbox{X-ray} emission to the SE
appears to trace the full extent of the lobe, and extends out to $\sim 65\arcsec$, corresponding
to a projected distance of $\sim 250$~kpc.  The emission to the NW extends out a similar distance 
as to the SE ($\sim 55\arcsec \approx 210$ kpc), but does not trace the full extent of the FRII 
radio lobe.  The \mbox{X-ray} emission to the NW is also weaker than to the SE.  There are 
$129\sp{+24}\sb{-22}$, $0.5-4$~keV background-subtracted counts in a 50\degree\/ wedge to the SE, 
extending from 16\arcsec\/ to 63\arcsec\/ (marked on Figure~\ref{fig:images}, bottom right, in
green)\footnote{Note that the SE region used here differs slightly from that used by M06, as the
\mbox{X-ray} emission appears to be diametrically opposed to the NW emission in this deeper exposure.},
and $86\sp{+24}\sb{-21}$ counts in a similar wedge along the NW counter jet.  The contributions
from point sources have been excluded and the values are corrected for any subsequent loss of area.
Owing to the increased exposure time of the combined observations, and the subsequent increase in
counts, we are able to conduct a basic spectral analysis of the diffuse emission.  We extract spectra
from the SE and NW wedge regions and bin them to a minimum of 15 counts per bin using \texttt{grppha}.
A power-law model gives $\Gamma=1.65\sp{+0.89}\sb{-0.74}$, 
$\chi\sp{2}/\nu=19.6/13$ and a flux of $F\sb{\rm 0.5-4\,keV}=(1.02\sp{+0.26}\sb{-0.25})\times10\sp{-14}\flux$ 
($L\sb{\rm 0.5-4\,keV}=1.81\sp{+0.46}\sb{-0.44}\times10\sp{42}\,\lum$) for the SE emission and 
$\Gamma=1.95\sp{+1.67}\sb{-1.05}$, $\chi\sp{2}/\nu=9.9/10$ and $F\sb{\rm 0.5-4\,keV}=(4.82\sp{+2.55}\sb{-1.80})\times10\sp{-15}\flux$
($L\sb{\rm 0.5-4\,keV}=8.56\sp{+4.5}\sb{-3.2}\times10\sp{41}\,\lum$) for the NW emission.  This
confirms the lower flux of the NW emission, and gives consistent power-law slopes.  Although we
cannot formally reject a bremsstrahlung model (it gives similar reduced \chisq values and
${\rm H0}>1\%$), the $kT$ value obtained is unconstrained, tending to higher temperatures and thus 
a power-law distribution.

Figure~\ref{fig:radial} (bottom) shows the radial surface-brightness profiles of the diffuse
\mbox{X-ray} emission.  The numbers of counts were extracted from 15 concentric arcs with equal
thickness between a radius of $0-60$\arcsec.  The profile of the SE diffuse emission (shown in
black) was extracted from a 50\degree\/ arc between angles 95\degree\/ and 145\degree, the
diffuse emission along the NW counterjet (shown in red) was extracted from a 50\degree\/ arc
between angles 275\degree\/ and 325\degree, and remaining portions of the arcs are used for 
the comparison profile (shown in green).  The background level (shown by the blue, dashed line)
was estimated from a $30\arcsec\times30\arcsec$ box at a distance $>80$\arcsec\/ from the central
source position.  This figure shows that the \mbox{X-ray} surface brightness along the jet and
counterjet directions is higher than that in the remaining directions.  The surface brightness
of the SE diffuse emission is also generally higher than that of the NW emission, although not
significantly at each radius.  In general, both profiles show a smooth decrease with increasing
radial distance from the source.\\


\section{Discussion}
\label{section:disc}
\subsection{The Nature of the X-ray and UV Absorption}
\label{section:disc_abs}
The \mbox{X-ray} absorption commonly observed in BAL quasars may be due to ``shielding gas'' located
close to the central black hole which prevents the over-ionization of the BAL wind
(e.g.,~\citealt{murray95,proga00,gibson09,wu10}).  Quasar-wind simulations suggest this gas should be
variable on timescales of months-to-yrs (e.g.,~\citealt{sim10,sim12}).  \mbox{X-ray} absorption
variability has been observed in some BAL quasars such as PG~$2112+059$ which showed a dramatic increase
in column density over 3~yrs \citep{gallagher04}, and PG~$1126-041$, a mini-BAL quasar which showed
rapid variability on timescales of months-to-hours due to highly ionized absorbers \citep{giustini11}.  
However, in the sample of 11 BAL quasars studied systematically by \citet{saez12}, exceptional \mbox{X-ray} 
variability (e.g.,~uncovering events where the shielding gas largely moves out of the line-of-sight),
compared to non-BAL quasars, was not seen.

In this work we have shown that the \mbox{X-ray} absorption properties of \pgten\/ measured in 4
observations taken up to 8.8~yrs (rest-frame) apart do not differ substantially, regardless of the
spectral model used.  When a fully covering and neutral absorber is assumed, the column densities
range from $N\sb{\rm H}=8\times10\sp{20}-4\times10\sp{21}$~\nhshort, which are among the lowest
values typically observed \citep{gallagher02,giustini08,fan09,streblyanska10}.  However, the 
power-law slopes obtained in the modeling are $\Gamma\le1.5$, perhaps indicating that the true 
levels of absorption in the source may be higher by a factor of $\sim2$ (see Section~\ref{section:fitting}).
In the case of a partially covering absorber, the column densities obtained are naturally higher,
$N\sb{\rm H}\sim10\sp{22}$~\nhshort, and more consistent with those for other BAL quasars.  The 
\mbox{X-ray} absorption can be well modeled with either a partially covering, or an ionized, absorber 
(see Table~\ref{table:fits}) and the spectral parameters of these models ($N\sb{\rm H}$, $f$, and 
$\xi$) are also consistent for each observation.  

As the \mbox{X-ray} absorption properties do not appear to vary, the absorption is perhaps more 
likely to be due to material located further from the central source, rather than the typical 
shielding gas expected on small scales.  In the case that \pgten\/ is intrinsically X-ray weak, 
such a shield is not required to ensure launching of the BAL wind.  Alternatively, if \pgten\/ 
is intrinsically \mbox{X-ray} normal, the shielding gas is required to be highly Compton-thick
($N\sb{\rm H}\approx7\times10\sp{24}\,\textrm{cm}\sp{-2}$; \citealt{luo13}).  Absorption with column
densities this high would not be detectable in \chandra or \xmm spectra as all \mbox{X-rays} below
$\sim10$~keV would be absorbed.  In both scenarios the \mbox{X-ray} emission we observe is therefore
predominantly from the jet; either because the continuum source is weak in comparison, or because
we only see \mbox{X-rays} from the portion of the jet extending beyond the shielding gas.  This
\mbox{X-ray} emission may then be absorbed by material further from the black hole, naturally
explaining the lack of variability and lower column densities observed.  However, we cannot rule
out variability occurring at a level below that which we are able to significantly detect with our
spectra.  A column density increase of more than a factor of 2 would be detected as a $3\sigma$
change between the \chandra spectra.

In the scenario in which \pgten\/ is intrinsically X-ray normal and is absorbed by Compton-thick 
shielding gas, a strong Fe \ka emission line at 6.4~keV with an EW of $1000-2000$~eV is expected
(e.g.,~\citealt{ghisellini94,matt96}).  Although we do measure a low level of neutral iron emission 
in a joint fitting of all 4 X-ray spectra (${\rm EW\sb{6.4~keV}=90\pm40~{\rm eV}}$;~see 
Section~\ref{section:fitting}), it is not as strong as might be expected.  This may be due to the 
dilution of the spectrum by jet-linked X-ray emission \citep{luo13}.  

The \mbox{X-ray} light curve in Section~\ref{section:lightcurve} shows significant flux variability
between the 4 \mbox{X-ray} observations.  Given the constant level of \mbox{X-ray} absorption present,
these changes are more likely related to intrinsic variations within the continuum source itself (in
the case that \pgten\/ is intrinsically \mbox{X-ray} weak), or fluctuations in the jet.

\begin{figure}
  \centering 
  \vspace*{0.2cm}
    \includegraphics[width=0.48\textwidth]{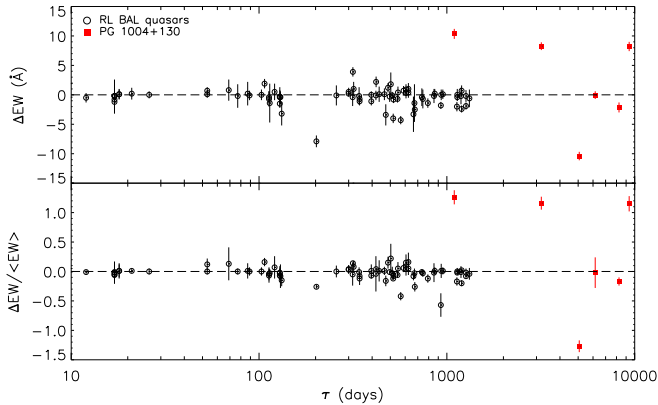}
  \caption{The \civ\/ BAL variability in \pgten, and other RL BAL
    quasars.  Filled red squares show the values for \pgten\/ and open
    black circles show values from the sample of RL BAL quasars
    presented in \citet{welling14} --- excluding \pgten.  The top
    panel shows the relationship between the absolute variability and
    rest-frame time interval.  The bottom panel shows the fractional
    variability.\\}
  \label{fig:bal_compare}
\end{figure}

In stark contrast to the \mbox{X-ray} absorption properties, the \civ\/ BAL in \pgten\/ shows
large-amplitude variability.  The archival observations we compare with our newly obtained \hst
COS spectrum probe rest-frame timescales of \mbox{$\approx~3-26$~yrs} during which large absolute and
fractional changes in EW are observed (see Table~\ref{table:uv_bal}).  However, the largest
changes are between the 2003 and 1986 spectra, i.e.,~$\Delta EW=-10.4\pm0.72$~\AA\/ and 
$\Delta EW/ \langle EW \rangle=-1.27\pm0.10$.  The magnitudes of these changes place \pgten\/ 
among the most variable RL and RQ BAL quasars (e.g.,~\citealt{filizak13}).  Our measurements 
probe long rest-frame timescales, on which BAL variability is observed to be stronger (see 
Section~\ref{section:intro}).  In Figure~\ref{fig:bal_compare} we compare the absolute and 
fractional BAL variability of \pgten\/ (red, solid points) to the sample of RL BAL quasars 
presented in \citeauthor{welling14}~(\citeyear{welling14};~open black circles).  This highlights 
the large variability of \pgten\/ in relation to other RL BAL quasars, even despite the longer 
rest-frame timescales probed by our observations. 

Variability in BAL troughs is likely caused by 2 main effects; clumps of material in the outflow moving
across our line-of-sight to the continuum source, or the material responding to changes in the incident
ionizing radiation.  When only portions of the BAL trough spanning a narrow velocity range show changes, perhaps
with some portions strengthening while others weaken (e.g.,~\citealt{gibson08,capellupo12}), clouds at
different velocities, and hence distances from the continuum source, entering and leaving the line-of-sight
is a plausible explanation, although it could also be a result of velocity-dependent optical depths 
\citep{capellupo12}.  However, many observations of BALs show co-ordinated changes in separate troughs of the same ion, 
for example, distinct \civ\/ BAL troughs tend to strengthen or weaken together \citep{filizak13}.  In this case, 
variations occurring over such a large velocity range are unlikely to be the result of co-ordinated 
movements within the outflow which would disperse the structure (e.g.,~\citealt{rogerson11}).  Instead, 
these, and global changes in which the entire profile of a single wide BAL trough varies uniformly 
(e.g.,~\citealt{grier15}), are likely a result of a change in the ionizing EUV continuum.  However this 
can only occur if the absorption trough is not too highly saturated.  Portions of the trough with low 
optical depths may show larger variations as they are more susceptible to changes in ionization, than 
those with only moderate optical depths.  

The changes observed in the BAL profile of \pgten\/ suggest a combination of these 
scenarios.  This is most apparent when considering the BAL trough in the 1986 observation, which
appears to be comprised of a main trough centered on $\lambda\sb{\rm rest}\sim1511$~\AA\/ ($v\sb{\rm out}\sim7830~\kms$) 
and an additional absorption feature with a smaller blueshift superimposed.  This component, at 
$\lambda\sb{\rm rest}\sim1526$~\AA\/ ($v\sb{\rm out}\sim4870~\kms$), does not appear in the other 3 observations (there 
is a hint of a similar component in the 1982 spectrum but its depth is within 10\% of the 
continuum level and could therefore be noise), and its narrow width suggests it may be due 
to cloud of material in the line-of-sight during the observation.  Making the assumptions that such 
a cloud is spherical, moving perpendicular to the line-of-sight and passing in front of a negligible 
size source, we can place an upper limit on the radius of the cloud of 
$r\lesssim1.3\times10\sp{17}~{\rm cm}\lesssim0.04~{\rm pc}$, as it must have traveled 
at least its diameter, at $v\sim4870~\kms$, in the 6163~day period.

The remainder of the BAL trough in the 1986 spectrum has a similar shape to that in 1982, 
although both the width and depth have increased by more than a factor of 2.  Such a 
monolithic change is more likely to be the result of the entire outflow varying in response to
a change in the ionizing continuum.  However, the UV flux does not change significantly between 
these observations.  Similarly, there is no significant flux difference between the 1986 and 
2003 observations, despite a large decrease in the BAL strength, $\Delta EW=-10.40\pm0.72$~\AA. 
The flux decrease between the 2003 and 2014 observations is only significant at $2\sigma$ but
does correspond to a strengthening of the BAL ($\Delta EW=8.26\pm0.66$~\AA).  The correlation 
between the observed UV flux and BAL EW may not be as strong as expected as the flux is measured 
over $\lambda\sb{\rm rest}=1411-1471$~\AA.  While this should give an indication of the UV flux 
variability, it is not directly measuring the unobservable EUV range which would be responsible 
for any changes within the outflow.  Previous observations have shown the EUV flux can vary more 
than the UV flux at longer wavelengths (e.g.,~\citealt{marshall97}).

If a change in the ionizing continuum is the main contributor to the BAL variability, we might 
expect to see co-ordinated variations in the mini-BAL.  This feature is centered on 
$\lambda\sb{\rm rest}\sim1542$~\AA\/ corresponding to an outflow velocity of $v\sb{\rm out}\sim1745~\kms$
which is over $10000~\kms$ slower than parts of the BAL trough.  Figure~\ref{fig:bal} shows that 
the mini-BAL in the 2003 \hst spectrum is significantly weaker than in the other 3 spectra.  This 
corresponds well to the BAL which is also extremely weak in this spectrum.  However, in general, this
feature shows much less variability than the BAL; in particular there is no corresponding deepening
in 1986 or 2014 when the BAL is considerably stronger.  This is likely due to the feature being 
more highly saturated, and therefore less susceptible to changes in the ionizing continuum, but 
may also suggest that material is indeed physically distinct from the higher velocity BAL and 
likely located much at a much greater distance from the central source. 

The 2014 BAL trough shows a similar shape to the main BAL trough in the 1986
spectrum, but is offset in wavelength.  This could be due to an acceleration of the entire outflow 
of $a\approx0.11\,{\rm cm\,s}\sp{-2}$, which is consistent with accelerations observed in other BAL 
quasars (e.g.,~\citealt{vilkoviskij01,rupke02,gibson08}).

In the standard model for BAL quasars the wind properties should depend on the \mbox{X-ray}/UV SED 
and the properties of the shielding gas.  Indeed, \civ\/ BALs are observed to be stronger and to have 
higher velocities in quasars with greater levels of X-ray absorption indicating that the shielding 
gas plays an important role in determining the properties of the outflowing wind 
\citep{brandt00,laor02,gallagher06,gibson09,wu10}.  In \pgten\/ the amount of \mbox{X-ray} absorption 
does not vary significantly with time, but despite this, large differences in the BAL EW are created.  
This further suggests that the \mbox{X-ray} absorption we observe is not due to the typical shielding 
gas.  The X-ray absorption could be due to the BAL outflow itself.  The presence of the \civ\/ ion in 
the UV material requires an ionization parameter in the range
\logxi$\sim-0.3\:{\rm to}\:+1.4$ (e.g.,~\citealt{kallman82}).  This is consistent with the values 
observed when the X-ray absorption is modeled with a partially ionized model (see Table~\ref{table:fits}).

We are unable to test for any relation between the \mbox{X-ray} 
flux and the EW of the UV BAL as only 2 UV and \mbox{X-ray} observations were made at similar epochs.  
However, despite large differences in EW, the \mbox{X-ray} flux is similar in each case, suggesting 
no relationship.  This is perhaps not unexpected as the observed \mbox{X-ray} emission in \pgten\/ is 
likely mostly from a small-scale jet (either because the continuum source is \mbox{X-ray} weak, or 
the \mbox{X-ray} normal continuum source is absorbed by Compton-thick shielding gas).  This flux may 
not be expected to influence the BAL production strongly.\\


\subsection{Nature of the Extra-nuclear X-ray Emission}
\label{section:disc_diff}
\mbox{X-ray} emission from extended jets has now been detected in over 100 AGN,\footnote{\url{http://hea-www.harvard.edu/XJET/}} 
but the physical processes involved in its production are still often uncertain.  \pgten\/ shows such
\mbox{X-ray} emission, a knot, in the same direction as the SE FRI radio jet; i.e.,~at a PA of 130\degree\/
located at a distance of 8\arcsec\/ (30~kpc) from the core.  No \mbox{X-rays} are detected in relation
to the NW FRII radio counterjet, but this is not unusual; the majority of detected \mbox{X-ray} jets are
one-sided, with the \mbox{X-rays} corresponding to the brighter of the radio jets \citep{worrall09}.
The HYMOR radio nature is often interpreted as a difference in the environment on either side of the
quasar; i.e.,~the jet propagating into a denser medium to the SE is quickly decollimated giving the FRI
appearance, whereas the jet propagating to the NW through a more tenuous medium appears as an FRII
(e.g.,~\citealt{gopal00}).  Shocks, created when the jet collides with an obstacle (e.g.,~\citealt{blandford79}),
accelerate electrons which then cool by emitting synchrotron \mbox{X-rays} and appear as bright \mbox{X-ray}
knots.  This scenario naturally explains why knots are commonly observed in FRI jets \citep{worrall09} 
as the probability of such shocks occurring is greater due to the higher density of the medium.  

Some \mbox{X-ray} knots have shown exceptional variability, such as the knot in the jet of M87 which
increased in intensity by 50\% over 5~yrs \citep{harris06a}.  However, no evidence for variability was
observed in the \mbox{X-ray} knots in Cen~A \citep{goodger10}, or in the quasar 3C~273 \citep{jester06}.
Although the knot in \pgten\/ does not show substantial variability between the 2 \chandra observations,
the $1.6\sigma$ decrease in flux (a factor of 2 in $\sim7.5$ rest-frame years) is consistent with electrons
that have been excited in a shock and are cooling via synchrotron emission, with typical loss-times for
such electrons being $\sim$ tens of yrs.  Expected power-law spectral indices for this synchrotron emission
are $\Gamma\sim1.6$ \citep{achterberg01}, which steepen to $\Gamma\sim2.1$ when energy losses become more
important.  Although we do not see a significant difference in the power-law slope between observations,
largely owing to the low number of counts in the spectra and hence large errors on the parameter, the
values we observe are broadly consistent with this scenario.

However, the \mbox{X-ray} knot is spatially extended by 4\arcsec\/ (15~kpc).  Since the loss-times of the
electrons are too short for them to diffuse across this distance before emitting \mbox{X-rays}, the emission
is likely being produced from a combination of multiple shock sites.  This may explain why the flux decrease
(and \gmm increase) is not highly significant, as we are observing the combined properties of these sites.

We note that in alternative models for the \mbox{X-ray} emission such as beamed IC/CMB \citep{tavecchio00,celotti01},
no flux drop is expected as the typical loss-times for the lower-energy electrons involved are $\sim10\sp{6}$~yrs. 
This scenario is also disfavored as the jet in \pgten\/ is likely inclined at $\theta\gtrsim45\degree$ (M06)
implying little line-of-sight beaming.

The peak of the \mbox{X-ray} knot emission lies closer to the core than the peak of the radio jet emission.
Offsets such as this are also commonly observed in other sources (e.g.,~\citealt{harris06b}), and are often
explained as a single population of electrons cooling by emitting radiation at progressively lower frequencies
from \mbox{X-ray} to radio as they propagate downstream along the jet \citep{bai03,hardcastle03}.  In such a
scenario optical emission may be expected between the \mbox{X-ray} and radio peaks, but this is not observed
in \hst imaging of \pgten\/ \citep{bahcall97,miller06}.  This perhaps indicates that the \mbox{X-ray} and radio
emission instead originate from different electron populations.

The deeper combined exposure time of both \chandra observations allows us to further constrain the nature of 
the diffuse \mbox{X-ray} emission which was previously observed in M06.  Figure~\ref{fig:images} (bottom right)
shows some of this emission extending to a radial distance of 65\arcsec\/ (250~kpc) from the core to the SE,
underlying the same area in which the broad FRI radio emission is observed, which includes both the jet and
the lobe.  Similar emission is observed to the NW, although it only extends to 55\arcsec\/ (210~kpc) and hence
does not trace the location of the FRII lobe which is located further from the quasar.  This suggests that the
diffuse \mbox{X-ray} emission in \pgten\/ is more closely related to the radio jet emission, rather than the
lobes.  The possible detection of a radio counterjet on small scales (C.~C.~Cheung, 2014, private communication)
also agrees with this interpretation.

\mbox{X-ray} spectra of the diffuse emission, extracted from the green wedges shown on Figure ~\ref{fig:images}
(bottom right), are well-fit with a power law.  This suggests that non-thermal processes are responsible for
the diffuse emission, a likely origin being unbeamed IC/CMB.  The diffuse \mbox{X-ray} emission is stronger to 
the SE ($L\sb{\rm 0.5-4\,keV}=1.81\sp{+0.46}\sb{-0.44}\times10\sp{42}\,\lum$) than the NW 
($L\sb{\rm 0.5-4\,keV}=8.56\sp{+4.5}\sb{-3.2}\times10\sp{41}\,\lum$).  This is consistent with the distribution 
of colder gas implied by the HYMOR nature of \pgten, which is likely responsible for the stronger radio 
jet to the SE and the weaker counterjet to the NW.  Electrons produced in the IC/CMB model are
of a relatively low energy, and hence have lifetimes of $\sim10\sp{6}$~yrs allowing them to travel large distances
before emitting \mbox{X-rays}.  The diffuse emission could therefore have an origin related to the jet and counterjet, 
but be located surrounding the radio jet as observed.\\


\section{Summary}
\label{section:sum}
We have presented new \textit{Chandra}, \textit{XMM-Newton}, and \hst COS observations of 
the RL BAL quasar \pgten.  We have compared these to archival \mbox{X-ray} and UV data in 
order to investigate potential changes in the \mbox{X-ray} and UV absorption, and in the 
\mbox{X-ray} jet.  We also combine both \chandra observations creating a deeper exposure 
with which to investigate the nature of the X-ray jet and diffuse X-ray emission in more 
detail than previously possible.  We summarize our main results below.

\begin{enumerate}
\item The amount of intrinsic \mbox{X-ray} absorption present in \pgten\/ does not appear 
      to vary significantly between each of the 4 \chandra and \xmm observations.  In each 
      case, the absorption can be well modeled with a partially covering absorber, or a 
      partially ionized absorber (best-fitting spectral parameters are listed in 
      Table~\ref{table:fits}).  The low column densities, 
      $N\sb{\rm H}=8\times10\sp{20}-4\times10\sp{21}$~\nhshort\/ for a fully covering 
      absorber, and lack of variability suggest that the absorption may not be due to the 
      typical shielding gas invoked in quasar-wind models.  See 
      Sections~\ref{section:fitting} and~\ref{section:disc_abs}.

\item The \mbox{X-ray} flux of \pgten\/ varies by up to $\sim40\%$ over 612~days (observed 
      frame; see Figure~\ref{fig:lc}), but remains at least a factor of $15\times$ weaker 
      than is expected for a typical RL quasar with a similar UV and radio luminosity.  
      As the source does not show significant variability in the amount of \mbox{X-ray} 
      absorption present, these changes are more likely related to intrinsic variations 
      within the continuum source itself, or fluctuations in the jet.  See 
      Sections~\ref{section:lightcurve} and~\ref{section:disc_abs}.
      
\item A strong \civ\/ BAL is observed in our new 2014 \hst COS observation, with $EW=11.24\pm0.56$~\AA.
      This is significantly variable, with an absolute change of $\Delta EW=8.26\pm0.66$~\AA\/ 
      and a fractional change of $\Delta EW / \langle EW \rangle=1.16\pm0.11$ from the previous 
      2003 \hst STIS observation, 3183~days earlier in the rest-frame of \pgten.  While the 
      absolute change in EW is not exceptional, the fractional change is one of the highest 
      observed in a BAL quasar.  See Sections~\ref{section:uv} and~\ref{section:disc_abs}.

\item An \mbox{X-ray} knot in the FRI radio jet is clearly detected in the combined, $\sim100$~ks 
      \chandra data (see Figure~\ref{fig:images}, top panels).  It appears downstream from 
      the peak of the radio emission, and at a distance of $\sim8\arcsec$ (30~kpc) from 
      the central \mbox{X-ray} source.  It has a spatial extent of $\sim4\arcsec$ (15~kpc).  
      Its spatial and spectral properties are consistent with synchrotron emission from 
      populations of shock-heated electrons.  No similar \mbox{X-ray} counterpart to the 
      FRII counterjet is detected.  See Sections~\ref{section:jet} and~\ref{section:disc_diff}.

\item Asymmetric diffuse \mbox{X-ray} emission is detected around \pgten, with a lower 
      luminosity underlying the unseen counterjet to the NW 
      ($L\sb{\rm 0.5-4\,keV}=8.56\sp{+4.5}\sb{-3.2}\times10\sp{41}\,\lum$) than the \mbox{X-ray} 
      detected jet to the SE ($L\sb{\rm 0.5-4\,keV}=1.81\sp{+0.46}\sb{-0.44}\times10\sp{42}\,\lum$).  
      In both cases a thermal model for the emission is disfavored and a non-thermal origin, 
      possibly due to inverse Compton scattering of CMB photons, is more likely.  The SE emission 
      has a best-fitting power-law slope of $\Gamma=1.65\sp{+0.89}\sb{-0.74}$, consistent with 
      the value for the NW emission, $\Gamma=1.95\sp{+1.67}\sb{-1.05}$.  In both cases, the 
      intensity of the emission decreases relatively smoothly, radially from the central source 
      (see Figure~\ref{fig:radial}, bottom).  It extends out to 65\arcsec\/ (250~kpc) to the SE 
      and 55\arcsec\/ (210~kpc) to the NW.  See Sections~\ref{section:diffuse} and~\ref{section:disc_diff}.

\end{enumerate}

In this work we have conducted a comprehensive, high S/N, multi-epoch, \mbox{X-ray} and UV spectral
monitoring campaign of the RL BAL quasar \pgten.  The results determined here serve as a template
for other RL BAL quasars.  The logical next step would be to study systematically
the properties of a larger sample of RL BALs (thus extending both this work and that of
\citealt{miller09}, \citealt{saez12}, and \citealt{welling14}).  However, the majority of such  
objects are not economical \chandra targets due to their high $z$ and low X-ray fluxes.  For example, 
to obtain 2 spectra with $\sim 500$ counts for each of the 12 snapshot targets from \citet{miller09} 
would require $>2$~Ms of \chandra time.  The forthcoming \textit{Athena} observatory, with its 
larger effective area, will provide a better facility for this type of study.


\acknowledgments 
AES, WNB, and BL gratefully acknowledge the support of \chandra \mbox{X-ray} Center 
grant G04-15093X, Space Telescope Science Institute grant HST-GO-13516.001-A, and 
NASA ADP grant NNX10AC99G.  SCG thanks the Natural Science and Engineering Research
Council of Canada for support.  The Guaranteed Time Observations (GTO) included here 
were selected by the ACIS Instrument Principal Investigator, Gordon P. Garmire, 
currently of the Huntingdon Institute for \mbox{X-ray} Astronomy, LLC, which is under 
contract to the Smithsonian Astrophysical Observatory; contract SV2-82024.  This work 
was also based on observations obtained with \textit{XMM-Newton}, an ESA science 
mission with instruments and contributions directly funded by ESA Member States and 
NASA.  We thank Teddy Cheung for sharing radio data with us, and the anonymous 
referee for constructive comments.\\

{\it Facilities:} \facility{CXO (ACIS)}, \facility{XMM (EPIC)}, \facility{HST (COS)}.


\bibliographystyle{apj}


\end{document}